\documentclass[final,1p,times]{elsarticle}

\usepackage{amssymb}
\usepackage{amsmath}
\usepackage{multirow}

\usepackage{lineno}

\journal{Icarus}
\date{September 2, 2024}

\begin{document}

\begin{frontmatter}



\title{Double ellipsoidal harmonic gravity field models for bilobed bodies: Example of comet 67P/Churyumov-Gerasimenko}


\author[inst1]{Xuanyu Hu}

\affiliation[inst1]{organization={Institute of Space Technology \& Space Applications, University of the Bundeswehr Munich},
            addressline={Werner-Heisenberg-Weg 39}, 
            city={Neubiberg},
            postcode={85577}, 
            country={Germany}}

\author[inst1]{Thomas P. Andert}

\begin{abstract}
Bilobed bodies represent a significant class of small extraterrestrial objects in the Solar System.
We present a double harmonic-series approach to model the gravity of the lobes separately, thereby allowing their mass distributions to be constrained independently.
We study an exemplary candidate contact binary, comet 67P/Churyumov-Gerasimenko, and establish two ellipsoidal harmonic series referred to the closest-bounding, non-overlapping ellipsoids of the respective lobes. 
We show that the model is more robust near the irregular body shape than a global spherical harmonic series, which is a customary product determined from spacecraft tracking.
While they can be estimated directly from measurements, we demonstrate here that the double series can also be transformed from an existing spherical harmonic model by means of decomposition.
We use simulations to analyze model solutions for a few heterogeneous mass distributions.
\end{abstract}




\end{frontmatter}


\section{Introduction}

Spacecraft missions have encountered several bilobed small bodies, which include comets \citep{2015Sci...347a1044S}, asteroids \citep{2006Sci...312.1330F,2013NatSR...3E3411H}, a trans-neptunian object 486958 Arrokoth \citep{stern2019initial}, and most recently, the moonlet of a binary asteroid Dinkinesch~$|$ Selam \citep{levison2024contact}, alluding to a substantial population at large \citep{2015aste.book..355M}.
The shape may indicate separate origins of individual objects before their merger in the aftermath of a nondestructive collision or accretion \citep{massironi2015two,jutzi2017formation,mckinnon2020solar}.

The knowledge of the body's gravity provides an important constraint on the interior mass distribution.
The determination of the gravity field model is a central task in most space missions, accomplished using observation of the perturbed spacecraft motion around the target, whenever such perturbations are detectable \citep{MG2000,TAPLEY2004xi}.
The model is often based on some harmonic series expansion. 
Since it is determined solely from the measured field variations with no assumption on the generating mass, the resulting series is a ``gravity'' model in the strict sense.
The series coefficients reveal directly certain linear combinations of the density moments, including such fundamental parameters as total mass, center-of-mass offset, and oblateness of the body mass.

Another class of methods evaluates the gravitation of a known or assumed mass distribution, such as a concentration of discrete mass elements, the density variations within the triangulated polyhedral mesh or the analytic function of the closed body surface \citep{paul1974gravity,pohanka1998optimum,balmino1994gravitational,russell2012global,1996CeMDA..65..313W,chen2019spherical}, etc.
This is often known as ``forward'' modeling.
The local, interior densities can be parameterized and estimated from observation or existing harmonic models, in which case the body shape and the surface morphology often provide the first hint (i.e., constraint) on the heterogeneity \citep{park2010estimating,tricarico2013global,takahashi2014morphology}.
Different distributions can produce the same exterior gravitational field.
Regardless of the formulation, the gravitational field models are theoretically indistinguishable (towards infinite resolution) in their respective regions of validity.

\subsection{Bi-harmonic approach}
When it comes to gravity modeling of bilobed bodies, it is natural and scientifically justified to treat the lobes separately to discern any potential difference between their interior structures.
Provided that separation is feasible, two lobe-dependent harmonic series indicate the permissible density moments of the respective distributions; collectively they must conform to the global gravitational field (model).
The approach was proposed and validated by \cite{andert2015gravity} for the nucleus of comet 67P/Churyumov-Gerasimenko, where the bi-harmonic model was formulated in a global, bipolar coordinate system with the foci coincident with the centers of the lobes \citep{jeffery1912form}.
The coefficients at degree zero indicate respectively the sum of and the difference between the lobes' bulk densities, in apparent analogy to the physical meaning of the spherical, and other, harmonic models.
The lobes' gravity was approximated by that of their respective best-fitting ellipsoids, as opposed to of two point mass \citep{zeng2015study,burov2019approximation}.
The approximations account for the dominant field components due to nonsphericity and are commonly adopted in astrodynamic analysis \citep{scheeres1994dynamics,wei2020hybrid}.


\subsection{Problem statement}
A triaxial ellipsoid is the appropriate reference surface for most of the convex, namely, single-lobed, objects.
The ``reference'' here does not imply an (ellipsoidal) approximation of the gravity field as mentioned above.
Rather, it is a geodetic surface to which the gravity field model is referred.
The model is an ellipsoidal harmonic (EH) series that theoretically can encompass infinite resolution \citep{2001CeMDA..79..235R}.
The coefficients measure specific combinations of the density moments of the body with respect to the reference ellipsoid, and can be derived from spherical harmonic coefficients via an exact transformation \citep{2016CeMDA.125..195H}.

Extending the investigation of \cite{andert2015gravity}, we employ double EH series for 67P, referred to the close(st)-bounding ellipsoids of the respective lobes in their local coordinate systems.
The reference ellipsoids are non-overlapping and yield an unambiguous partition of the mass.
The model is valid or absolutely convergent in the free space exterior to the reference ellipsoids.

The EH coefficients can be then used in the same way as in a global model to interpret or constrain the interior distribution of each lobe.
They can be estimated from observation, e.g., tracking of spacecraft motion \citep{2002GeoRL..29.1231G,park2014gravity}.
Here we show and substantiate that the double EH model can be derived from an existing global model, such as the spherical harmonic (SH) series.
We investigate via simulation the feasibility of the model application to distinguish between several candidate cases of inhomogeneity.

\section{Gravitational field models and coordinate systems}

\subsection{Spherical and ellipsoidal harmonic models}
We assume that a ``global'' model is available in the most common form of a SH series. The gravitational potential is thereby formulated as:

\begin{equation} \label{eq:sh}
V = \displaystyle \sum^{N_\mathrm{max}}_{N=0} \sum^{N}_{M=0} \left( \frac{r_0}{r} \right)^{N+1} P_{NM}( \sin \varphi ) \left( C_{NM} \cos M\lambda + S_{NM} \sin M\lambda \right),
\end{equation}
where $\lambda,\varphi,r$ are the spherical coordinates of longitude, latitude, and radius of a field point. $r_0$ is the reference radius. $P_{NM}$ is the \textit{normalized} associated Legendre polynomial of degree $N$ and order $M$ (where we omit the common symbol of an overhead bar indicating normalization to declutter the notation). $C_{NM},S_{NM}$ are the coefficients determined by the body's gravitational field, or simply field coefficients. Note that they have the same dimension as potential, i.e., $G\mathcal{M}/r_0$ with $\mathcal{M}$ being the total body mass, which differs from a more common, dimensionless convention (e.g., $C_{00}=1$). The maximum degree, $N_\mathrm{max}$, specifies the model resolution.

The SHs form a set of orthogonal functions on a sphere:

\begin{align} \label{eq:orth_sh}
\int^{2\pi}_0 \int^{\pi/2}_{-\pi/2} P_{NM}&(\sin\varphi) \left[\begin{matrix}
\cos M\lambda \\ \sin M\lambda    
\end{matrix}\right] P_{N'M'}(\sin\varphi) \left[\begin{matrix}
\cos M'\lambda \\ \sin M'\lambda    
\end{matrix}\right] \cos\varphi\mathrm{d}\varphi\,\mathrm{d}\lambda \nonumber \\ &= \left\{ \begin{matrix} 4\pi, & \text{if}\ N=N', M=M' \\ 0, & \text{otherwise.} \end{matrix} \right.
\end{align}
If the potential is known over the reference sphere, the orthogonality can be used as a band-pass filter allowing the field coefficients at given frequencies to be extracted via a boundary value problem.\\

The potential can be alternatively expressed as an EH series,

\begin{equation} \label{eq:eh}
V = \displaystyle \sum^{n_\mathrm{max}}_{n=0} \sum^{2n+1}_{m=1} V_{nm}, \quad
V_{nm} = c_{nm}\  \frac{F_{nm}(\rho)}{F_{nm}(\rho_0)} E_{nm}(\mu) E_{nm}(\nu),
\end{equation}
which satisfies the Laplace's equation in terms of the ellipsoidal coordinates of $\rho$, $\mu$, and $\nu$, found, in the descending order, as the square root of the real roots of the cubic equation in terms of $s$, $x^2/s + y^2/(s-h^2)+ z^2/(s-k^2)=1$, for known Cartesian coordinates and some real, positive constants $k>h$. The ellipsoidal coordinates thus specify a system of three quadratic, orthogonal surfaces, an ellipsoid, a one-sheet and a two-sheet hyperboloid, respectively, all of focal lengths $h,k$ and intersecting at $(\pm x,\pm y,\pm z)$. $E_{nm}$ and $F_{nm}$ are the Lam{\'e} polynomials of the first and second kind, respectively, of degree $n$ and order $m$. The latter plays the same role as the attenuation factor, $r^{-N-1}$, in Eq. (\ref{eq:sh}). $\rho_0$ denotes the semimajor axis of the reference ellipsoid. $c_{nm}$ are the field coefficients.

The orthogonality of the normalized EHs refers to the ellipsoid of focal lengths $h$ and $k$,

\begin{equation} \label{eq:orth_eh}
\iint_S E_{nm}(\mu)E_{nm}(\nu)E_{n'm'}(\mu)E_{n'm'}(\nu) \mathrm{d}S = \left\{ \begin{matrix} \pi/2, & \text{if}\ n=m', m=m' \\ 0, & \text{otherwise,} \end{matrix} \right.
\end{equation}
where 

\begin{equation}
\iint_S \mathrm{d}S = \int^h_0 \int^k_h \frac{(\mu^2-\nu^2) \mathrm{d}\mu \mathrm{d}\nu}{\sqrt{(\mu^2-h^2)(k^2-\mu^2)(h^2-\nu^2)(k^2-\nu^2)}} = \frac{\pi}{2},
\end{equation}
is the solid angle of an ellipsoidal octant.

\subsection{Division of shape and definition of lobes}
\begin{figure}
    \centering
    \includegraphics[width=\linewidth]{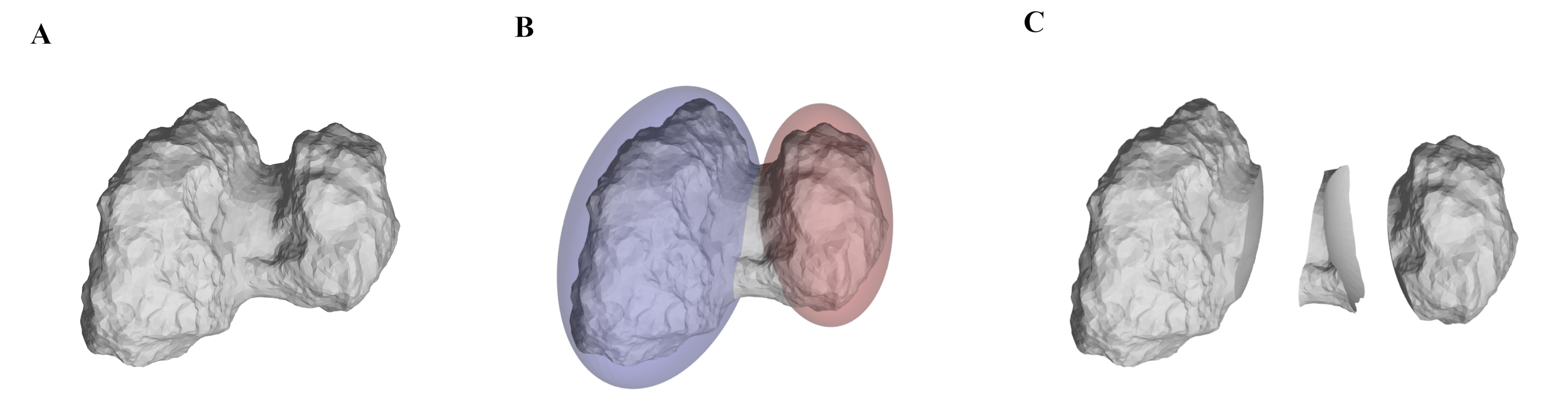}
    \caption{Shape model and partition of 67P.}
    \label{fig:67p-3p}
\end{figure}

The shape model of 67P is as shown in Fig. \ref{fig:67p-3p}A \citep{2016Icar..277..257J}. Two planes, $0.787x-0.545y+0.27z=0.29$ and $-0.919x-0.291y+0.267z=-1$, given in the same study, divide the shape into a large and a small lobe, sometimes vividly referred to as the ``body'' and the ``head'', respectively, and a connecting ``neck'' of the rubber duck-shaped nucleus.
The reference ellipsoids of the lobes were found as the smallest enclosing ellipsoids of the vertices of the respective polyhedra and do not overlap (Fig. \ref{fig:67p-3p}B).
They (re)partition the global shape model into three polyhedra: the vertices inside the large and the small ellipsoids belong to the respective lobes, which we number Lobe 1 (``body'') and Lobe 2 (``head''); those outside both ellipsoids belong to the neck (Fig. \ref{fig:67p-3p}B).
Note that the resulting polyhedra comprise at least one (signed) surface segment of the reference ellipsoid cut along the path of intersection to ensure closure (Fig. \ref{fig:67p-3p}c). In other words, an ellipsoidal surface segment is shared by, and encloses, the lobe and the neck (polyhedra) with opposite surface normals.

The semiaxes of the reference ellipsoid for the large lobe are 2.334, 1.877, 1.2~km (red in Fig. \ref{fig:67p-3p}B); those for the small lobe are 1.585, 1.348, 0.9371~km (blue in Fig. \ref{fig:67p-3p}B). The corresponding focal lengths are given in Table \ref{tab:css}.

\subsection{Local coordinate systems and notation} \label{ss:loccord}
The local model given by Eq. (\ref{eq:eh}), written above in a generic form, must be expressed in a lobe-dependent coordinate system.
Let $\mathbf{r}^\mathrm{T} = (x\ y\ z)$ denote a position vector and its Cartesian coordinates in the global system. The coordinate transformation into the local system associated with Lobe $i$ is

\begin{equation} \label{eq:cstran}
\mathbf{r}^{(i)} = \mathbf{R}_{i} \left( \mathbf{r} - \mathbf{t}_{i} \right),
\end{equation}
where $\mathbf{t}_i$ points to the origin of the local coordinate system and where $\mathbf{R}_{i}$ is a $3 \times 3$ rotation matrix (Table \ref{tab:css}).
Accordingly, we distinguish the quantities associated with the local system by the superscript ``($i$)''.
For example, $V^{(i)}$ is the potential of Lobe $i$ evaluated by Eq. (\ref{eq:eh}) with field coefficients $c^{(i)}_{nm}$ derived in the local coordinate system.\\


\begin{table}[]
    \centering
    \caption{Parameters of reference ellipsoids, translation vectors and rotation matrices from global into local coordinate systems} \label{tab:css}
    \begin{tabular}{c l|c|c}
    \hline
    \hline
       &  & Lobe 1 (``Body'') & Lobe 2 (``Head'') \\
    \hline
    $\rho^{(i)}_0, h^{(i)}, k^{(i)}$ & (km) & 2.334, 1.387, 2.002 & 1.585, 0.835, 1.279 \\
    \hline
        $\mathbf{t}_i$ & (m) & $(-737.3\ \ 166.8\ -142.0)^\mathrm{T}$ & $(1553\ -370.1\ \ 133.6)^\mathrm{T}$ \\
    \hline
        $\mathbf{R}_i$ & & $\left(\begin{matrix} 0.6927 & 0.706 & -0.1476 \\ 0.2137 &-0.3964 &-0.8929 \\-0.6888 & 0.5869 &-0.4255 \end{matrix}\right)$ & $\left(\begin{matrix} 0.2344 & 0.902 & 0.3625 \\-0.3887 &-0.2548 & 0.8854 \\0.891 &-0.3484 & 0.2909 \end{matrix}\right)$ \\
    \hline
    \end{tabular}
\end{table}

\section{SH and double EH model for 67P}
We start with a basic simulation to demonstrate the feasibility of the double EH approach and test the consistency of the model implementation.
The model performance is also compared to that of the SH model.
The gravitation of a polyhedron is evaluated via the forward method by \citep{1996CeMDA..65..313W}, which transforms the volume integration into line integrals along the (facet) edges.
The calculation is exact for the adopted density in the simulation setting.
The potential on the reference surface provides then the boundary condition to solve for the field coefficients.

\subsection{Derivation of field coefficients}
With the potential on the reference sphere, $V(r_0,\varphi,\lambda)$, the orthogonality of Eq. (\ref{eq:orth_sh}) enables the SH coefficients to be derived as follows \citep{HM1967}:

\begin{align}
\left[\begin{matrix} C_{NM}\\ S_{NM} \end{matrix}\right] &= \frac{1}{4\pi} \int^{2\pi}_0 \int^{\pi/2}_{-\pi/2} V P_{NM}(\sin\varphi) \left[\begin{matrix}
\cos M\lambda \\ \sin M\lambda
\end{matrix}\right] \cos\varphi\mathrm{d}\varphi \mathrm{d}\lambda \nonumber \\
&= \frac{1}{4\pi} \int^{2\pi}_0 \int^{\pi/2}_{-\pi/2} V_{NM} P_{NM}(\sin\varphi) \left[\begin{matrix}
\cos M\lambda \\ \sin M\lambda
\end{matrix}\right] \cos\varphi\mathrm{d}\varphi \mathrm{d}\lambda\,.
\end{align}
The calculation of the EH coefficients of Lobe $i$ (i.e., distinguishing between two local coordinate systems) follows Eq. (\ref{eq:orth_eh}),

\begin{equation}
c^{(i)}_{nm} = \frac{1}{4\pi}\iint_{S^{(i)}}\, V^{(i)} E^{(i)}_{nm}(\mu^{(i)})E^{(i)}_{nm}(\nu^{(i)})\, \mathrm{d}S^{(i)},
\end{equation}
with $V^{(i)}$ over the reference ellipsoid of semimajor axis $\rho^{(i)}_0$.

We reemphasize that appropriate coordinate transformations by Eq. (\ref{eq:cstran}) must precede the application of the models at a field point in global coordinates.

\subsection{Case study: homogeneous nucleus}
\begin{figure}
    \centering
    \includegraphics[width=\linewidth]{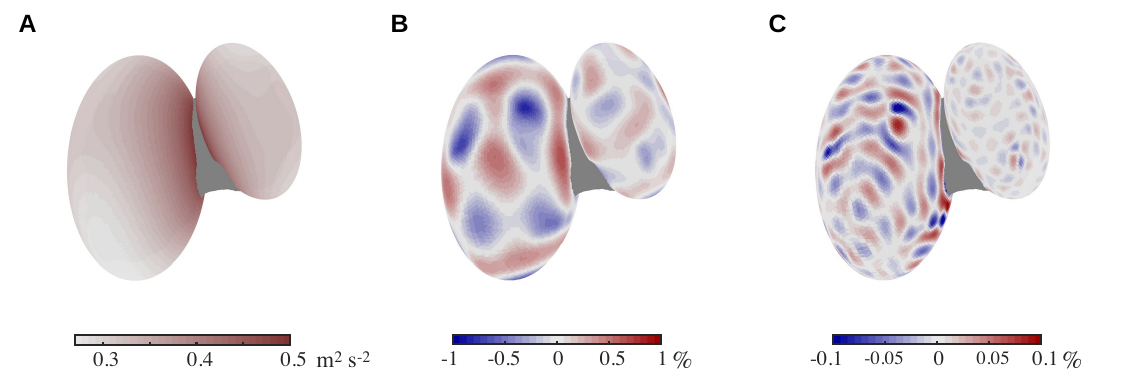}
    \caption{Potential and double-EH model errors on reference ellipsoids. \textbf{A}. Potential due to two lobes. \textbf{B}. Evaluation errors of double EH model up to degree 5. \textbf{C}. Evaluation errors up to degree 15, where the colormap scale is reduced by a factor of 10 compared with that in \textbf{B}.}
    \label{fig:deh-err_re}
\end{figure}

We assume that each lobe has the same, constant density of 530~$\mathrm{kg/m^3}$ \citep{patzold2016homogeneous,patzold2019nucleus}.
The gravitational potential over the reference ellipsoids is depicted in Fig. \ref{fig:deh-err_re}A.
The areas inside the polyhedral shape were excluded.
The evaluation errors of a model are defined as

\begin{equation} \label{eq:def_modelerr}
\Delta V = \frac{ \hat{V}-V }{ V } \times 100 \%
\end{equation}
which is the difference between the model evaluated potential, $\hat{V}$, and the (true) polyhedral potential, $V$.

The errors exhibit even undulations between -0.83 and 0.98~\% with an RMS of 0.68 \% for EH models up to degree 5 (Fig. \ref{fig:deh-err_re}B). 
They are reduced, as expected, to the range between -0.13 and 0.11~\% and the RMS of 0.053~\% when the model resolutions are increased to degree 15 (panel C). 
The reduction as well as the higher-frequency pattern indicate the errors result from the truncation of the series leading to signal loss at higher, omitted frequencies. They are thus effectively alleviated by incorporating higher-degree and -order harmonics.\\

The evaluation of the surface gravitation within the reference ellipsoid risks the validity of the harmonic series and thereby numerical stability of the model.
Nonetheless, the reference ellipsoids are (almost) the closest reference surfaces to the bi-lobed shape, it is of practical interest to examine the model errors on the body surface, or at least, the extent to which they can still be considered tolerable.

\begin{figure}
    \centering
    \includegraphics[width=\linewidth]{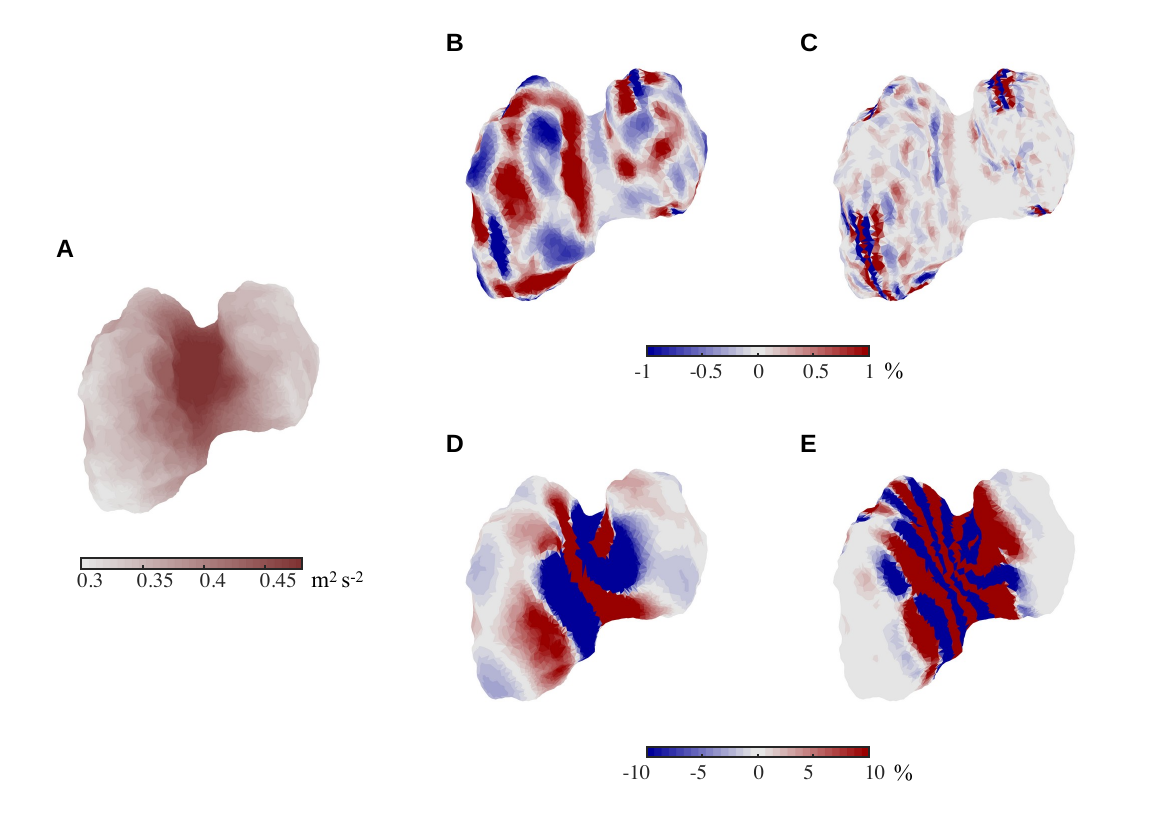}
    \caption{Potential and model errors on the nucleus surface of 67P. \textbf{A}. Potential over the surface of 67P. \textbf{B}. Errors of double EH model up to degree 5. \textbf{C}. Errors of double EH model up to degree 15. \textbf{D}. Errors of SH model up to degree 5. \textbf{E}. Errors of SH model up to degree 15. Out-of-range errors are uniformly represented by colors at the respective limits.}
    \label{fig:deh-err_srf}
\end{figure}

The variation of the surface potential is comparable to that on the reference ellipsoids in magnitude (Fig. \ref{fig:deh-err_srf}A), as a probable consequence of the proximity of the reference ellipsoids to the surface.
The model errors up to degree 5 are also similar in pattern (Fig. \ref{fig:deh-err_re}B) but span a greater range between -10.7 and 6.4 \%. 
Raising the model resolution to degree 15, we observe an  improvement over a majority of the surface. Other areas, e.g., around the tip of the small lobe and on the near side of the large lobe, are found with an intensification of errors to the absolute maximum of 200 \% far beyond the displayed color scale.
The regions of non-improvement, which account for about 20 \% of the surface area, are either concavities or abrupt departures from the overall triaxial shapes of the lobes, and situated further underneath the reference surface than their surroundings.
The errors growing with model resolution are divergent in character.

In fact, nothing illustrates better the divergence effect than the behavior of the SH model in this case (Fig. \ref{fig:deh-err_srf}D and E).
From a resolution of degree 5, the model is crippled by extreme errors reaching $2.3\times 10^4$~\% over the neck region and extending to the lobes overlooking it.
Increasing the maximum degree to 15 would cause further deterioration at higher frequencies, with the maximum error soaring to the (even more) meaningless $3 \times 10^{10}$~\% (panel E).
It would be biased to disregard the fact that the SH model can also be more reliable than the EH model, where the reference sphere lies closer to the body.
However, the advantage exists only locally, e.g., in the areas where the double EH model is notably divergent (panel C).
For instance, substantial (absolute) errors greater than 10~\% occur over about 42~\% of the surface area for the SH model and 1.4~\% in the case of the double EH model. 

The above results can be compared with various previous studies on the use of EH series for gravitational field modeling of small bodies \citep{2001CeMDA..79..235R,2015JGeod..89..159H}.
We note the high-degree global model for 67P was developed by \citep{2016JGRE..121..497R} using logarithmic expressions.

\section{Transformation of SH into double EH model}
The EH model coefficients can be estimated in the spacecraft orbit determination process.
In the case of double series, the local coordinate systems described in Section \ref{ss:loccord} are involved to evaluate the lobes' perturbations on the spacecraft in the inertial frame.
For validating purposes, we consider a simplified least-squares solution of deriving the EH models from a global SH series.
The problem retains the essence of local model implementation in connection to observables in the global frame.
\\

The decomposition of the global model into local entities is founded on the theoretical equivalence between the potential given by Eq. (\ref{eq:sh}) and the sum of contributions from all components,

\begin{equation}
\displaystyle V \equiv V^{(1)} + V^{(2)},
\end{equation}
assuming infinite model resolutions in Eqs. (\ref{eq:sh}) and (\ref{eq:eh}), respectively.
The potential due to the neck is neglected for the time being.\\

We recognize two basic (and obvious) conditions:

I) It is only feasible to transform the global model into local models with reduced resolution(s). The total number of SH field coefficients is $\mathcal{N}_\mathrm{SH} = (N_\mathrm{max}+1)^2$.
Let the number of coefficients in a local model be denoted accordingly as $\mathcal{N}^{(i)}_\mathrm{EH}=(n^{(i)}_\mathrm{max}+1)^2$. We have

\begin{equation} \label{eq:cond1}
\mathcal{N}^{(1)}_\mathrm{EH}+\mathcal{N}^{(2)}_\mathrm{EH} \le \mathcal{N}_\mathrm{SH},
\end{equation}
unless additional constraints are applied. Therefore, $n^{(i)}_\mathrm{max}<N_\mathrm{max}$.
The equality rarely occurs in Eq. (\ref{eq:cond1}).
The over-determined problem is expressed by the linear system:

\begin{equation} \label{eq:ls_obs}
\mathbf{y} = \mathbf{H} \mathbf{x} + \mathbf{e},
\end{equation}
where the observables are the known SH coefficients,

\begin{equation} \label{eq:obsvec}
\mathbf{y} = \left( C_{00}\ C_{10}\ C_{11}\ S_{11}\ \cdots\ C_{NM}\ S_{NM}\cdots\right)^\mathrm{T}.
\end{equation}
The vector of $(\mathcal{N}^{(1)}_\mathrm{EH}+\mathcal{N}^{(2)}_\mathrm{EH})$ unknown parameters is

\begin{equation} \label{eq:estvec}
\mathbf{x} = \left( c^{(1)}_{01}\ c^{(1)}_{11}\ c^{(1)}_{12}\ \cdots\ c^{(1)}_{nm}\ \cdots\ \mid\ c^{(2)}_{01}\cdots c^{(2)}_{nm}\ \cdots \right)^\mathrm{T},
\end{equation}
where coefficients for each lobe are grouped together. $\mathbf{e}$ is the error vector.

The observation matrix comprises column-wise two blocks:
\begin{equation} \label{eq:obseq}
\mathbf{H} = \left(\, \begin{matrix} \mathbf{H}^{(1)} & \mathbf{H}^{(2)} \end{matrix}\, \right),
\end{equation}
in which $\mathbf{H}^{(i)}$ is $\ \mathcal{N}_\mathrm{SH} \times \mathcal{N}^{(i)}_\mathrm{EH}\ $  in dimension.
Suppose that $c^{(i)}_{nm}$ is the $j$th estimate in $\mathbf{x}$ (Eq. \ref{eq:estvec}) and $C_{NM}$ or $S_{NM}$ the $l$th observable in $\mathbf{y}$ (Eq. \ref{eq:obsvec}). The element at the $l$th row and $j$th column of $\mathbf{H}^{(i)}$ can be computed via a boundary-value problem as

\begin{equation} \label{eq:obs_elem}
h^{(i)}_{lj} = \frac{1}{4\pi} \displaystyle \int^{2\pi}_0 \int^{\pi/2}_{-\pi/2} v^{(i)}_{nm}\, P_{NM}(\sin \varphi) \left[ \begin{matrix} \cos M\lambda \\ \sin M\lambda \end{matrix} \right] \, \cos\varphi \mathrm{d}\varphi \mathrm{d}\lambda
\end{equation}
where $v^{(i)}_{nm} = \left(F^{(i)}_{nm}(\rho^{(i)})/F^{(i)}_{nm}(\rho^{(i)}_0)\right) E^{(i)}_{nm}(\mu^{(i)})E^{(i)}_{nm}(\nu^{(i)}) = V^{(i)}_{nm}/c^{(i)}_{nm}$ is the solid EH (Eq. \ref{eq:eh}) evaluated over the reference sphere of radius $r_0$, discretized into a grid of 200$\times$100 points in longitude and latitude, respectively.

The system is solved as:

\begin{equation} \label{eq:ls}
\hat{ \mathbf{x} } = \mathbf{P} \mathbf{H}^\mathrm{T} \mathbf{W}\, \mathbf{y},\quad \mathbf{P} = \left( \mathbf{H}^\mathrm{T} \mathbf{W} \mathbf{H}\right)^{-1},
\end{equation}
which minimizes $\hat{\mathbf{e}}^\mathrm{T} \mathbf{W} \hat{\mathbf{e}}$, where $\mathbf{W}$ is the weight matrix and $\hat{\mathbf{e}}=\mathbf{y}-\mathbf{H}\hat{\mathbf{x}}$ the residual vector.
\\

II) The omission of coefficients beyond $n_\mathrm{max}$ causes the estimated coefficients, $\hat{c}_{nm}$, to become erroneous.
The omission is embodied in the error vector, $\mathbf{e}$, in Eq. (\ref{eq:ls_obs}).
As the amplitude of the coefficients decreases with frequency, the impact will be exacerbated at lower resolution.
Thus, the issue can be most acute when decomposing the SH model into double EH series where a significant reduction of resolution is necessary.
Note that a SH model is susceptible to truncation errors, as well.
The topic is however irrelevant to the present study.

\subsection{Test case: From degree-1 SH to double degree-0 EH models} \label{ss:test_case}
To appreciate the impact of the truncation, we consider the simplest case of $N_\mathrm{max}=1$ (or $\mathcal{N}_\mathrm{SH}=4$), where we derive two EH models up to degree 0, i.e., $c^{(1)}_{01}$ and $c^{(2)}_{01}$ or their gravitational parameters (GMs).
Let us set up the observables and the estimation vector as follows:

 \begin{equation}
\mathbf{y} = \left( \begin{matrix} C_{00} & C_{11} & S_{11} & C_{10} \end{matrix} \right)^\mathrm{T},\quad \mathbf{x} = \left( \begin{matrix} c^{(1)}_{01} & \mid & c^{(2)}_{01} \end{matrix} \right)^\mathrm{T}.
\end{equation}
Meanwhile, we define an augmented vector:

\begin{equation} \label{eq:augvec}
\tilde{\mathbf{x}} = \left( \begin{matrix} \mathbf{x} \\ \mathbf{c}\end{matrix} \right),\quad \mathbf{c} = \left(\begin{matrix} c^{(1)}_{11} & c^{(1)}_{12} & c^{(1)}_{13} & \mid & c^{(2)}_{11} & c^{(2)}_{12} & c^{(2)}_{13} \end{matrix} \right)^\mathrm{T},
\end{equation}
where $\mathbf{c}$ contains 6 additional, consider parameters of degree-1 EH coefficients for both lobes, which are \textit{not} solved for but whose impact will be assessed.
The observation equation becomes

\begin{equation} \label{eq:obs_aug_d1}
\mathbf{y} = \boldsymbol{\mathcal{H}} \left(\begin{matrix} \mathbf{x} \\ \mathbf{c} \end{matrix} \right), \quad \boldsymbol{\mathcal{H}} = \left( \begin{matrix}
\mathbf{H} & \mathbf{H}_\mathrm{c} \end{matrix} \right).
\end{equation}
Of course, the solution can only be obtained for $\hat{\mathbf{x}}$ via Eq. (\ref{eq:ls}) using $\mathbf{H}$, while the augmented system based on $\boldsymbol{\mathcal{H}}$ is underdetermined.

\subsubsection{Physical meaning of field coefficients}
The degree-0 and -1 SH coefficients according to the expression of Eq. (\ref{eq:sh}) are related respectively to the GM and Cartesian coordinates of the center of mass, say $\boldsymbol{\sigma} = (x_\sigma\ y_\sigma\ z_\sigma)^\mathrm{T}$,

\begin{equation} \label{eq:sh-com}
\quad C_{00} = \frac{G\mathcal{M}}{r_0};  \quad
\left( \begin{matrix} C_{11} \\ S_{11} \\ C_{10} \end{matrix} \right) = \frac{G\mathcal{M}}{\sqrt{3}(r_0)^2} 
\left( \begin{matrix} x_\sigma \\ y_\sigma \\ z_\sigma \end{matrix} \right).
\end{equation}
We are reminded that all coefficients are of the same dimension as $G\mathcal{M}/r_0$.
Should they be defined to be dimensionless, where $C_{00}=1$, the degree-1 coefficients would be multiplied by the common factor of $\sqrt{3}r_0$ to yield the center-of-mass coordinates.

The corresponding EH coefficients are likewise proportional to the $G\mathcal{M}_i$ and $\boldsymbol{\sigma}^{(i)}$ of each lobe:

\begin{align} \label{eq:eh-com}
&\quad c^{(i)}_{00} = G\mathcal{M}_i\ F^{(i)}_{01}(\rho^{(i)}_0); \nonumber \\
&\left( \begin{matrix} c^{(i)}_{11} \\ c^{(i)}_{12} \\ c^{(i)}_{13} \end{matrix} \right) = G\mathcal{M}_i\, \mathbf{D}^{(i)} \left( \begin{matrix} x^{(i)}_\sigma \\  y^{(i)}_\sigma \\ z^{(i)}_\sigma \end{matrix} \right), \quad \mathbf{D}^{(i)} =\left( \begin{matrix} d^{(i)}_x & 0 & 0 \\ 0 & d^{(i)}_y & 0 \\ 0 & 0 & d^{(i)}_z \end{matrix} \right).
\end{align}
 Due to triaxiality of the ellipsoidal system, however, $c_{11},c_{12},c_{13}$ are \textit{not} in uniform scale to the center-of-mass coordinates, i.e., the diagonal elements in $\mathbf{D}$ are distinct (\ref{se:app1}), as opposed to the SH with a single reference length of $r_0$ and consequently a constant coefficient-to-coordinate ratio of $G\mathcal{M}/(\sqrt{3}r^2_0)$.\\
 
\subsubsection{Centers of mass of two lobes}
\begin{figure}
    \centering
    \includegraphics[width = \linewidth]{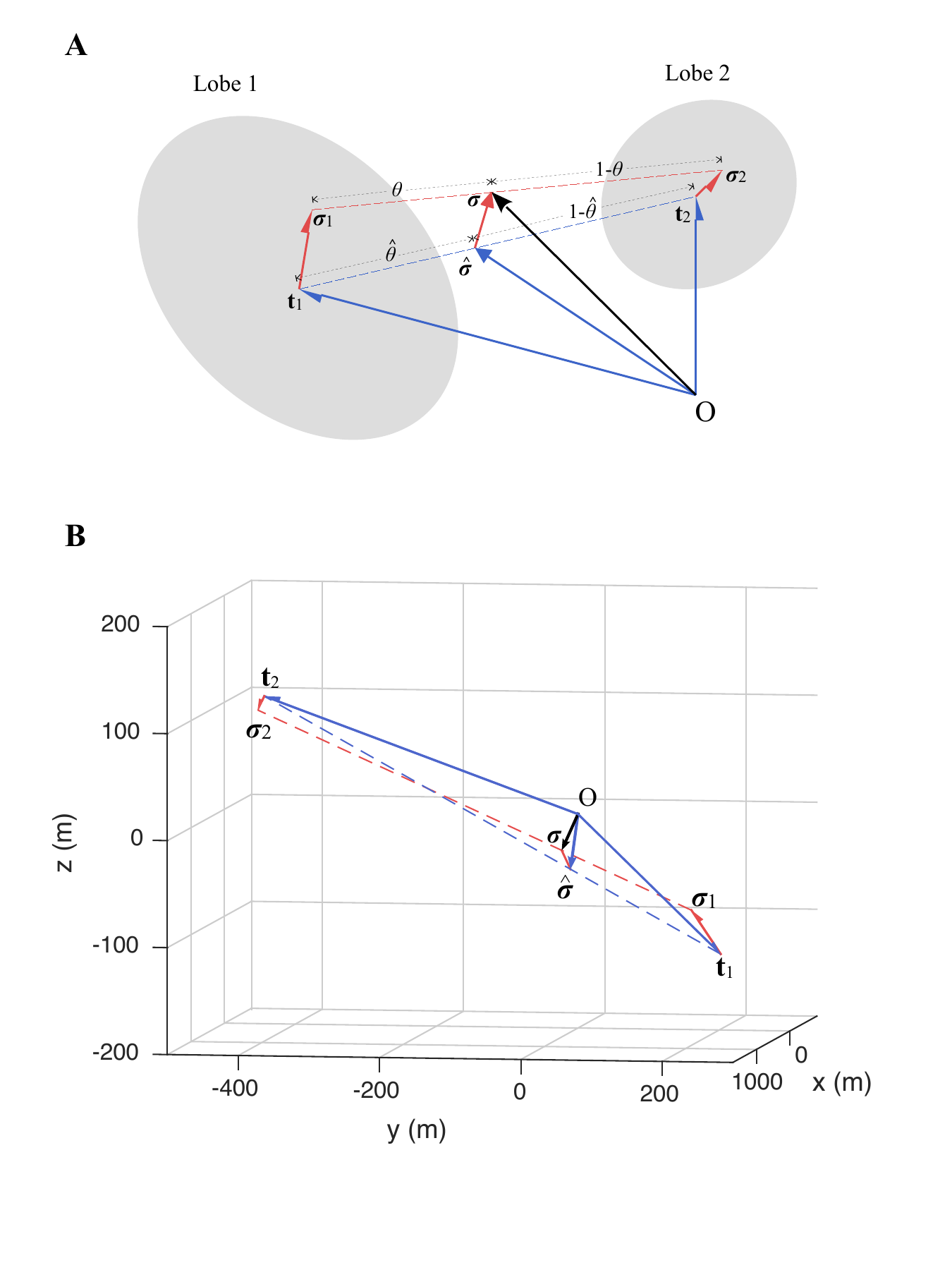}
    \caption{Schematic illustration of the global center of mass relative to the lobes' (\textbf{A}) according to Eq. (\ref{eq:com-coms_simple}) and visualization of the actual least-squares solution (\textbf{B}). The shorthand symbols are $\theta=\mathcal{M}_2/\mathcal{M}$ and $\hat{\theta}=\hat{\mathcal{M}}_2/\hat{\mathcal{M}}$ (in \textbf{A}).}
    \label{fig:com}
\end{figure}

The center-of-mass position is related to the first-order mass moments of the lobes (while still excluding the neck),

\begin{align}
 \boldsymbol{\sigma} &=  \frac{1}{\mathcal{M}} \int_\mathrm{L1+L2} \mathbf{r}\,\mathrm{d}\mathcal{M} \nonumber \\
 &= \frac{1}{\mathcal{M}_1+\mathcal{M}_2} \left( \int_\mathrm{L1} \mathbf{r}\,\mathrm{d}\mathcal{M} +  \int_\mathrm{L2} \mathbf{r}\,\mathrm{d}\mathcal{M} \right),
\end{align}
where $\mathcal{M}_i=\int_{\mathrm{L}i} \mathrm{d} \mathcal{M}$ is the total mass of Lobe $i$.
Let the integrals be evaluated in the respective local coordinate systems, so that (recalling Eq. \ref{eq:cstran} for the reverse transformation),

\begin{equation} \label{eq:r_com2}
 \boldsymbol{\sigma} = \frac{1}{\mathcal{M}} \sum_i \int_{\mathrm{L}i} \left( \mathbf{R}^\mathrm{T}_i \mathbf{r}^{(i)} + \mathbf{t}_i \right) \,\mathrm{d}\mathcal{M}.
\end{equation}
Recognizing that the first part of the integral is related to the center-of-mass position of each lobe in their own coordinate system, which we denote by $\boldsymbol{\sigma}^{(i)}$, Eq. (\ref{eq:r_com2}) is rewritten as

\begin{equation} \label{eq:com-coms}
 \boldsymbol{\sigma} = \sum_i \frac{\mathcal{M}_i}{\mathcal{M}} \left(  \mathbf{R}^\mathrm{T}_i \boldsymbol{\sigma}^{(i)} + \mathbf{t}_i \right), \quad \boldsymbol{\sigma}^{(i)} = \frac{1}{\mathcal{M}_i} \int_{\mathrm{L}i} \mathbf{r}^{(i)} \mathrm{d}\mathcal{M}.
\end{equation}
For the purpose of clarity, we collect the expression in the parentheses as $\boldsymbol{\sigma}_i = \mathbf{R}^\mathrm{T}_i \boldsymbol{\sigma}^{(i)} + \mathbf{t}_i$, which transforms $\boldsymbol{\sigma}^{(i)}$ into the global coordinate system.
Then,

\begin{equation} \label{eq:com-coms_simple}
\boldsymbol{\sigma} = \sum_i \frac{\mathcal{M}_i}{\mathcal{M}} \boldsymbol{\sigma}_i = \frac{\mathcal{M}_1}{\mathcal{M}_1+\mathcal{M}_2} \boldsymbol{\sigma}_1 + \frac{\mathcal{M}_2}{\mathcal{M}_1+\mathcal{M}_2} \boldsymbol{\sigma}_2, 
\end{equation}
expresses basically the colinearity of the global and the lobes' centers of mass, as visualized in Fig. \ref{fig:com}A.
The condition will be used to test the consistency of the least-squares solution.

\subsubsection{Least-squares solution, validation, and implication}
\noindent\textbf{Observation matrix}\\
We substitute the expressions of the SH and EH coefficients in terms of $\boldsymbol{\sigma}$ and $\boldsymbol{\sigma}^{(i)}$ by Eqs (\ref{eq:sh-com}) and (\ref{eq:eh-com}) into the left- and right-hand sides of (\ref{eq:obs_aug_d1}) for $\mathbf{y}$ and $\tilde{\mathbf{x}}$, respectively. Comparing the resulting equation with Eq. (\ref{eq:com-coms}), we find that the augmented observation matrix, $\boldsymbol{\mathcal{H}}$, in Eq. (\ref{eq:obs_aug_d1}) should take the form 

 \begin{align} \label{eq:obs_analyt}
 &\mathbf{H} = \left( \begin{matrix} r_0 &  \\  & \sqrt{3}r^2_0\,\mathbf{I}_{3} \end{matrix} \right)^{-1}
 \left( \begin{matrix} 1 & 1 \\
  \mathbf{t}_1 &  \mathbf{t}_2 \end{matrix} \right)
 \left( \begin{matrix} F^{(1)}_{01}(\rho^{(1)}_0) &  \\  & F^{(2)}_{01}(\rho^{(2)}_0) \end{matrix} \right)^{-1}, \nonumber \\
 &\mathbf{H}_\mathrm{c} = \frac{1}{\sqrt{3} r^2_0}\left( \begin{matrix} \mathbf{0}_{1 \times 3} & \mathbf{0}_{1 \times 3} \\ \mathbf{R}_1 & \mathbf{R}_2 \end{matrix} \right) \left( \begin{matrix}
\mathbf{D}^{(1)} & \\ & \mathbf{D}^{(2)}    
 \end{matrix} \right)^{-1},
 \end{align}
where $\mathbf{I}_3$ is a 3$\times$3 identity matrix and where $\mathbf{R}_i, \mathbf{t}_i$ are as initially defined in Eq. (\ref{eq:cstran}); $\mathbf{D}^{(i)}$ is as in Eq. (\ref{eq:eh-com}).

The result differs from the numerically evaluated observation matrix via Eq. (\ref{eq:obs_elem}) by a maximum of $10^{-8}$, which is the expected numerical uncertainty of the quadrature (over a 200$\times$100-point grid).\\

\noindent\textbf{Error and interpretation}\\
The least-squares estimates, $\hat{\mathbf{x}}=( \hat{c}^{(1)}_{01}\ \hat{c}^{(2)}_{01})^\mathrm{T}$, the corresponding $G\hat{\mathcal{M}}_i$, and their errors are given in Table \ref{tab:gm_est}.
The errors are caused by neglecting $\mathbf{c}$ (in Eq. \ref{eq:augvec}) and can be quantified as
$\Delta \mathbf{x} = \mathbf{S}\,\mathbf{c}$, where $\mathbf{S} =-\mathbf{P}\,\mathbf{H}^\mathrm{T} \mathbf{W} \mathbf{H}_\mathrm{c}$ is the sensitivity matrix \citep{MG2000,TAPLEY2004xi}.
The residual vector, $\hat{\mathbf{e}}=\mathbf{y}-\mathbf{H} \hat{\mathbf{x}}$, contains the errors in the reconstructed SH field coefficients of degree 1 (Table \ref{tab:resvec}). 

On the other hand, the Eq. (\ref{eq:com-coms_simple}) offers a physical interpretation of the solution and the errors.
In principle, the global center of mass is colinear with, and lies between, those of the lobes, as depicted by the vectors, $\boldsymbol{\sigma}, \boldsymbol{\sigma}_1, \boldsymbol{\sigma}_2$, along the dashed red line in Fig. \ref{fig:com}A.
A practical solution of $\hat{\boldsymbol{\sigma}}_i$ is out of the question here due to under-determinacy (Eq. \ref{eq:obs_aug_d1}).
By neglecting the effect of $\boldsymbol{\sigma}_i$, the solution is performed along the line (segment) between $\mathbf{t}_1$ and $\mathbf{t}_2$, marked by the dashed blue line.
The actual solution is shown in Figure \ref{fig:com}B for comparison.

In effect, $\hat{c}^{(1)}_{01},\hat{c}^{(2)}_{01}$ are determined in such a way that they locate a point, $\hat{\boldsymbol{\sigma}}$, between $\mathbf{t}_1, \mathbf{t}_2$, which is the closest to the global center of mass, $\boldsymbol{\sigma}$.
We can thus formulate an equivalent problem as follows.
Let $\theta = \mathcal{M}_2/(\mathcal{M}_1+\mathcal{M}_2)$.
The position of a point between $\mathbf{t}_1$ and $\mathbf{t}_2$ is given by

\begin{equation} 
\mathbf{p} = \mathbf{t}_1 + \left( \mathbf{t}_2 - \mathbf{t}_1 \right) \theta,\quad \text{for}\ \theta \in [0,1].
\end{equation}
It is easy to show that the squared distance between $\mathbf{p}(\theta)$ and $\boldsymbol{\sigma}$ is

\begin{equation} \label{eq:min-dst-soln}
||\mathbf{p}-\boldsymbol{\sigma}||^2 = \underbrace{||\mathbf{t}_2 - \mathbf{t}_1 ||^2}_a\,\theta^2 + \underbrace{2\left(\mathbf{t}_2 - \mathbf{t}_1\right) \cdot \left(\mathbf{t}_1 - \boldsymbol{\sigma} \right)}_b \theta + ||\mathbf{t}_1 - \boldsymbol{\sigma}||^2.
\end{equation}
The quadratic equation of $||\mathbf{p}(\theta)-\boldsymbol{\sigma}||^2=0$ has no real roots.
The minimum of the function occurs at $\theta=-b/(2a)$= 0.32881 yielding $G\hat{\mathcal{M}}_2 = 214.62$~$\mathrm{m^3\,s^{-2}}$.

The result differs from the least-squares solution by 0.0076~$\mathrm{m^3\,s^{-2}}$ (which is rounded off in Table \ref{tab:gm_est}).
The reconstructed center-of-mass coordinates likewise approach the least-squares estimates, $\hat{\boldsymbol{\sigma}}$, within $(-0.4\ 0.9\ 0.5)^\mathrm{T}$~mm and yield indistinguishable residuals in Table \ref{tab:resvec}.
The reason for this marginal discrepancy is that, however slightly, the least-squares problem still differs from Eq. (\ref{eq:min-dst-soln}).
The latter has only one unknown, $\theta$, or $\mathcal{M}_2/\mathcal{M}$, and $\mathcal{M}_1/\mathcal{M}$ is always $1-\theta$.
In comparison, both $\mathcal{M}_1$ and $\mathcal{M}_2$ are estimated in the least-squares problem.
Note that we may also eliminate one of them similarly in the observation equation (\ref{eq:obseq}), thereby enforcing $G\hat{\mathcal{M}}_1+G \hat{\mathcal{M}}_2 \equiv G\mathcal{M}$.
The advantage of doing so is imperceptible, as is evidenced by the small difference (residual) of $0.0235$~$\mathrm{m^3\,s^{-2}}$ (or 0.0036~\%) between $G\hat{\mathcal{M}}$ and the actual (Table \ref{tab:gm_est}).
The improvement would thus be minimal.
We also confirm that such an approach would not affect the results involving higher-resolution models reported hereafter.\\

\begin{table}
    \centering
    \caption{Test case: Solution and errors of transformed double EH model} \label{tab:gm_est}
    \begin{tabular}{c c |c c c}
    \hline\hline
 Estimate: $\hat{\mathbf{x}}$ & ($\mathrm{m^2\,s^{-2}}$) & \multicolumn{2}{c}{Gravitational parameter} \&& error ($\mathrm{m^3\,s^{-2}}$)  \\
        \hline
  $\hat{c}^{(1)}_{01}$ & 0.24454 &\quad\quad $G\hat{\mathcal{M}}_1$ & 438.09 & -19.498 \\ 
  $\hat{c}^{(2)}_{01}$ & 0.16705 &\quad\quad $G\hat{\mathcal{M}}_2$ & 214.62 &  \ 19.475 \\  
       \hline
  Total & - &\quad\quad $G\hat{\mathcal{M}}$ & 652.71 &-0.0235 \\ 
  \hline
    \end{tabular}
\end{table}


\begin{table}
    \centering
    \caption{Test case: Residuals of reconstructed degree-1 SH field coefficients} \label{tab:resvec}
    \begin{tabular}{c c r|c r c}
    \hline\hline
  Residual: $\hat{\mathbf{e}}$& \multicolumn{2}{c|}{($\mathrm{m^2\,s^{-2}}$)} &  Center of mass: $\boldsymbol{\sigma}$ & (m)   \\
        \hline
   $\Delta C_{11}$ & & 2.4180$\times 10^{-4}$ & $\Delta x_\sigma$ & 5.072  \\
   $\Delta S_{11}$  & & 6.043$\times 10^{-4}$ & $\Delta y_\sigma$ & 12.674  \\
   $\Delta C_{10}$  & & -8.3192$\times 10^{-4}$ & $\Delta z_\sigma$ & -17.452  \\
       \hline
    \end{tabular}
\end{table}

\begin{table}
    \centering
    \caption{Test case: Residuals of reconstructed degree-1 SH field coefficients} \label{tab:resvec}
    \begin{tabular}{c | c c c c |c c c c }
    \hline\hline
   & \multicolumn{4}{c|}{Lobe 1 (``Body'')} & \multicolumn{4}{c}{Lobe 2 (``Head'')}   \\
    & $c^{(1)}_{01}$ & $c^{(1)}_{11}$ & $c^{(1)}_{12}$ & $c^{(1)}_{13}$ & $c^{(2)}_{01}$ & $c^{(2)}_{11}$ & $c^{(2)}_{12}$ & $c^{(2)}_{13}$  \\
   \hline
   ($10^{-3}\ \mathrm{m^2s^{-2}}$) & 257 & 4.46 & 0.226 &-10.9 & 150 & -0.25 & 1.34 & -2.86 \\
   error (\%) & 0.566 & 12.9 & 99.4 & 2.19 & 1.33 & 19.1 & 18.5 & 33.0 \\
       \hline
    \end{tabular}
\end{table}

To sum up of the test case, we have verified that the omission of the degree-1 EH coefficients of the lobes introduced an error, which could be fully traced and explained in the simulation.
The issue will occur in higher degrees and orders, so long as the omitted coefficients are non-negligible.
Loosely speaking, the coefficients, especially towards low degrees, are not random, white noises.
They contribute expressly to the SH coefficients at (from) the same degree. And, such a contribution cannot be eliminated by minimization of the difference with the SH model of higher resolution.



\subsection{Higher-degree model transformation}
We next analyze how the estimation errors are influenced by the choice of the model resolutions.
Because the SH and the EH models have different resolutions, i.e., $\mathcal{N}_\mathrm{SH}>\mathcal{N}^{(i)}_\mathrm{EH}$, we consider two scenarios accordingly.

\subsubsection{SH model resolution}
From the text case follows naturally the question, whether increasing the resolution of the global SH model can improve the solution of the degree-0 double EH model.
The reasoning is that raising the SH model resolution increases the number of observations, which can be effective, e.g., when $\mathbf{e}$ consists of only white noises. 

\begin{figure}
    \centering
    \includegraphics[width = \linewidth]{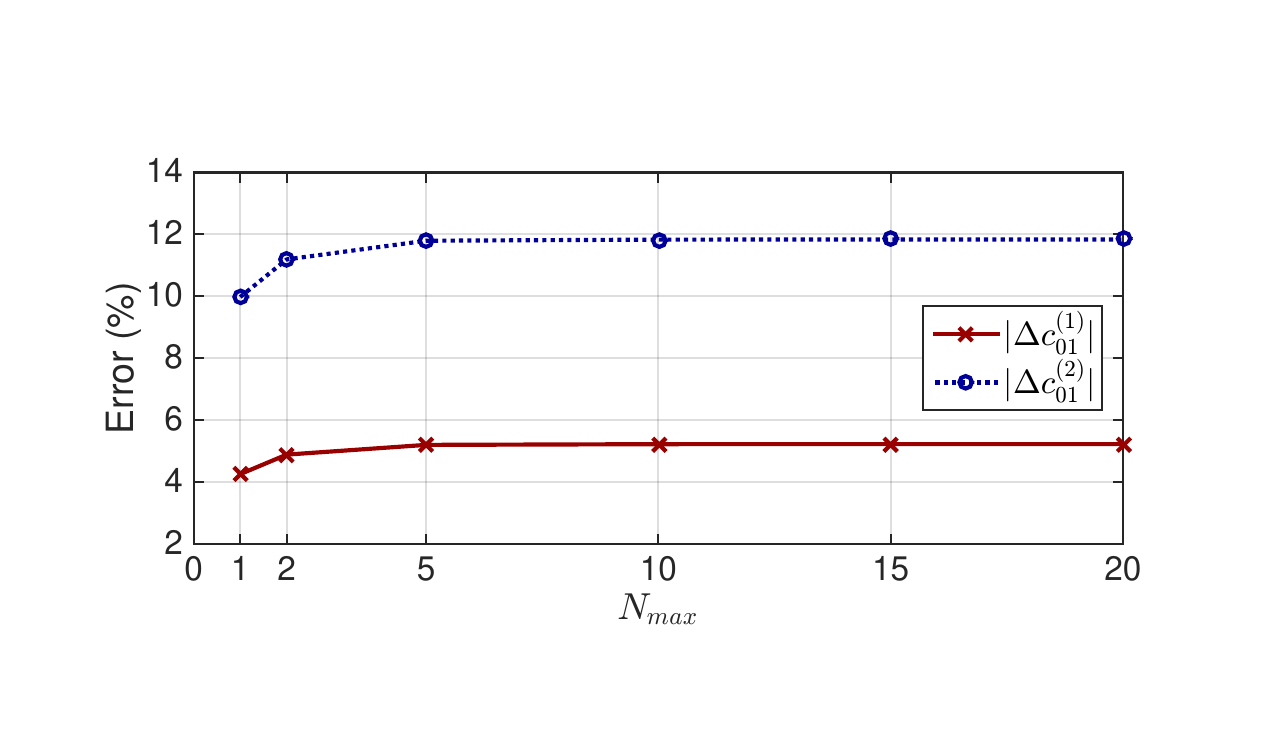}
    \caption{Absolute percentage errors of double EH degree-0 coefficients as function of original SH model resolution.} \label{fig:sh-hger-res}
\end{figure}

The EH model errors shown in Figure \ref{fig:sh-hger-res} disproves this impression.
We see that the errors grow with the maximum degree of the SH model and plateau from degree 5.
This bears out the conclusion from the test case, that the impact of omitted EH coefficients on the SH coefficients cannot be eliminated by residual minimization.
Enhancing the SH resolution while still keeping the minimum EH resolution implies that the omission effect of higher-degree EH harmonics would be somewhat intensified.
The curves eventually stabilized as the SH coefficients themselves tail off in magnitude towards higher degrees.

\subsubsection{EH model resolution(s)}
\begin{figure}
    \centering
    \includegraphics[width = \linewidth]{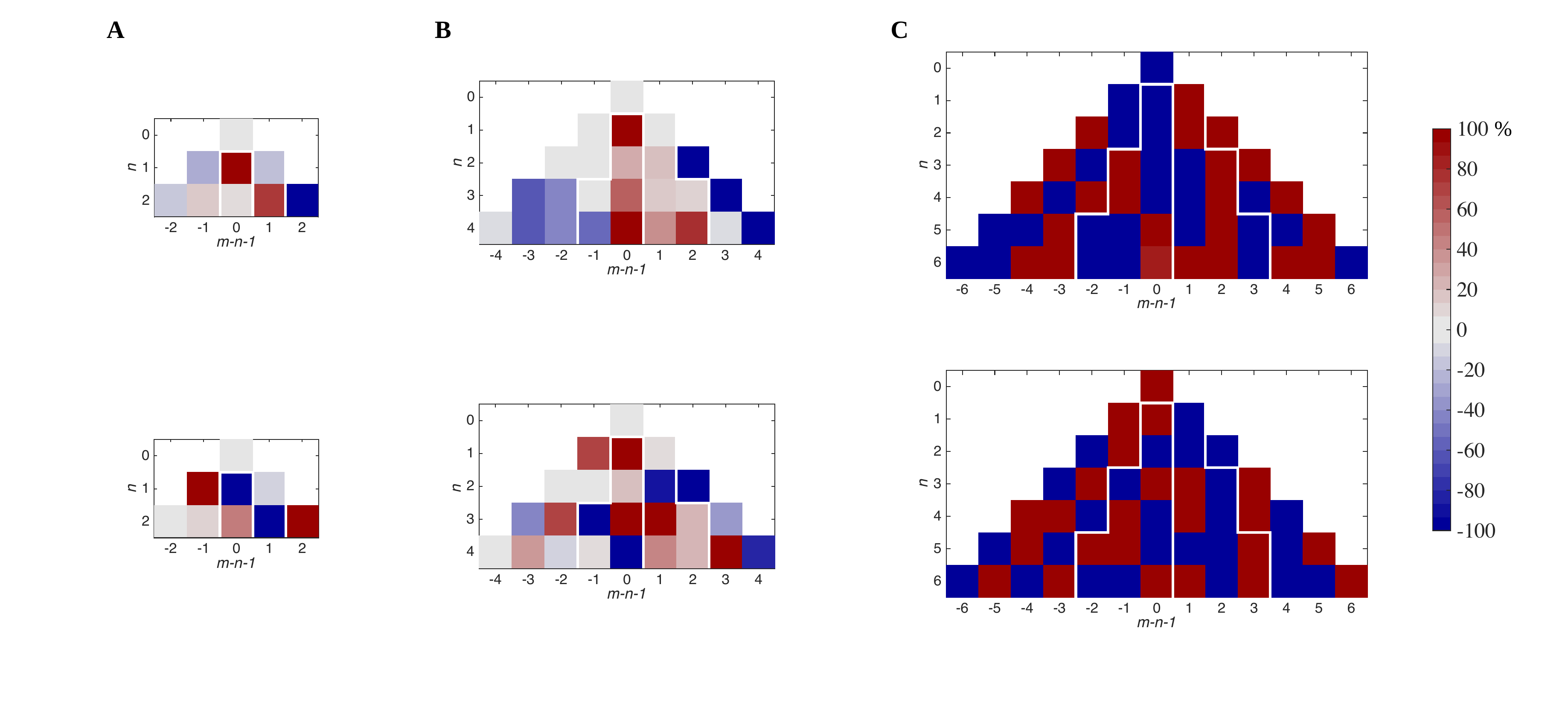}
    \caption{Percentage errors of field coefficients of double EH model up to degree 2 (\textbf{A}), 4 (\textbf{B}), and 6 (\textbf{C}). Results for Lobe 1 are in the first row, those for Lobe 2 in the second. The horizontal axis is defined by the symmetric order, $m-n-1$.} \label{fig:eh-hger-res}
\end{figure}

It may be thus beneficial to account for or estimate higher-degree EHs in the solution.
The characterization of the EH error behavior with changing model resolution is more complicated.
A reason is that the (maximum) resolution of the model depends on that of the base SH model (Eq. \ref{eq:cond1}), not to mention that the two EH models need not be of the same resolution.
We restrict ourselves here to the case $\mathcal{N}^{(1)}_\mathrm{EH}=\mathcal{N}^{(2)}_\mathrm{EH}$.
Figure \ref{fig:eh-hger-res} shows the errors in the coefficients of the double EH model up to degrees 2, 4, and 6, all based on a global degree-10 SH model.
At a glance (before a more quantitative analysis below), the degree-4 double EH model (panel B) is more accurate than the degree-2 (panel A), most notably at degree 1 and the first two harmonics of degree 2 ($m{-n}{-1}$=-2,-1 or $m$=1,2) for both lobes.
Note that a comparison beyond degree 2 is not possible.
Degree 6, with 98 EH coefficients, is the highest resolution that could be reached from 121 SH coefficients.
The solution is altogether compromised with substantial errors from 86 to $3.5\times10^5$~\%.
\\

\begin{figure}
    \centering
    \includegraphics[width = .8\linewidth]{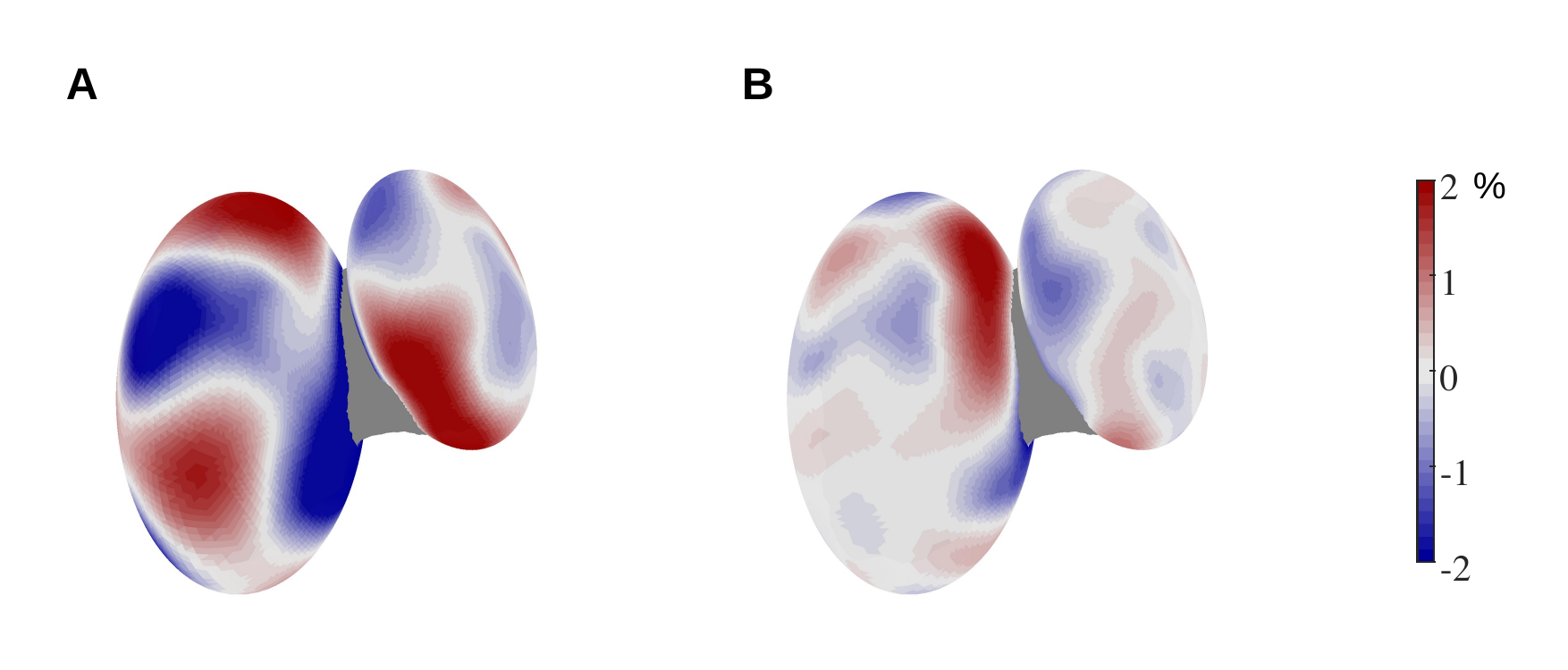}
    \caption{Errors of modeled potential on the reference ellipsoids up to degree 2 (\textbf{A}), 4 (\textbf{B}).} \label{fig:eh-hger-res-potl}
\end{figure}

It is more straightforward to compare the degree-2 and -4 models based on evaluation errors on the reference lobes.
The error of the degree-2 model ranges between -4.37 and 2.47~\%, with an RMS of 3.32~\%.
Indicative of the ``normal'' model behavior (see Fig. \ref{fig:deh-err_re}), the errors were alleviated in the degree-4 model, narrowed down to between -1.95 and 2.22~\% with an RMS of 1.46~\%, thus quantitatively supporting the observed improvement in Fig. \ref{fig:eh-hger-res-potl}B.
The pattern of increased frequency in addition suggests the errors within the model resolution have been effectively reduced and the remainder comprises omitted signals (Fig. \ref{fig:eh-hger-res}).\\


\section{Strategy for improvement: Model compensation}
We propose a method here to improve the double EH solution, namely, to remedy the errors of the coefficients to a certain extent.
The idea is simple: for the chosen EH model resolution, $n^{(i)}_\mathrm{max}$, we compensate the coefficients beyond this degree.
Perhaps somewhat ironically, this would necessitate assumptions on the density.
Thus, the treatment is strictly understood as a remedy, or a supplement to the double EH model.
We will, however, test the effectiveness or restriction of the remedy for some common interior structures.

\subsection{Bulk compensation of coefficients beyond maximum degree}
Let us refer to the augmented observation equation (\ref{eq:obs_aug_d1}), and note that the consider parameter therein, $\mathbf{c}$, accommodates the omitted coefficients for $n>n^{(i)}_\mathrm{max}$, which was specified as degree 1 in the test case.
With perfect knowledge of $c^{(i)}_{nm}$ beyond $n^{(i)}_\mathrm{max}$, we could compensate for the omission effect in the SH field coefficients.
We point out beforehand that this is \textit{impossible} in reality, since we can neither know nor make a perfect assumption about the mass distribution.
The most obvious approach is to assume the omitted coefficients can be attributed to a homogeneous interior with a constant, bulk density.
Then, the observables can be compensated as follows,

\begin{equation} \label{eq:obs_compensate}
 \tilde{\mathbf{y}} = \mathbf{y}-\mathbf{H}_\mathrm{c} \tilde{\mathbf{c}}, \quad \tilde{\mathbf{c}} = \left( \tilde{c}^{(1)}_{n^{(1)}_\mathrm{max}+1,1} \cdots\ |\ \tilde{c}^{(2)}_{n^{(2)}_\mathrm{max}+1,1} \cdots \right)^\mathrm{T},
\end{equation}
to replace $\mathbf{y}$ in Eq. (\ref{eq:ls}).\\

\begin{figure}
    \centering
    \includegraphics[width = \linewidth]{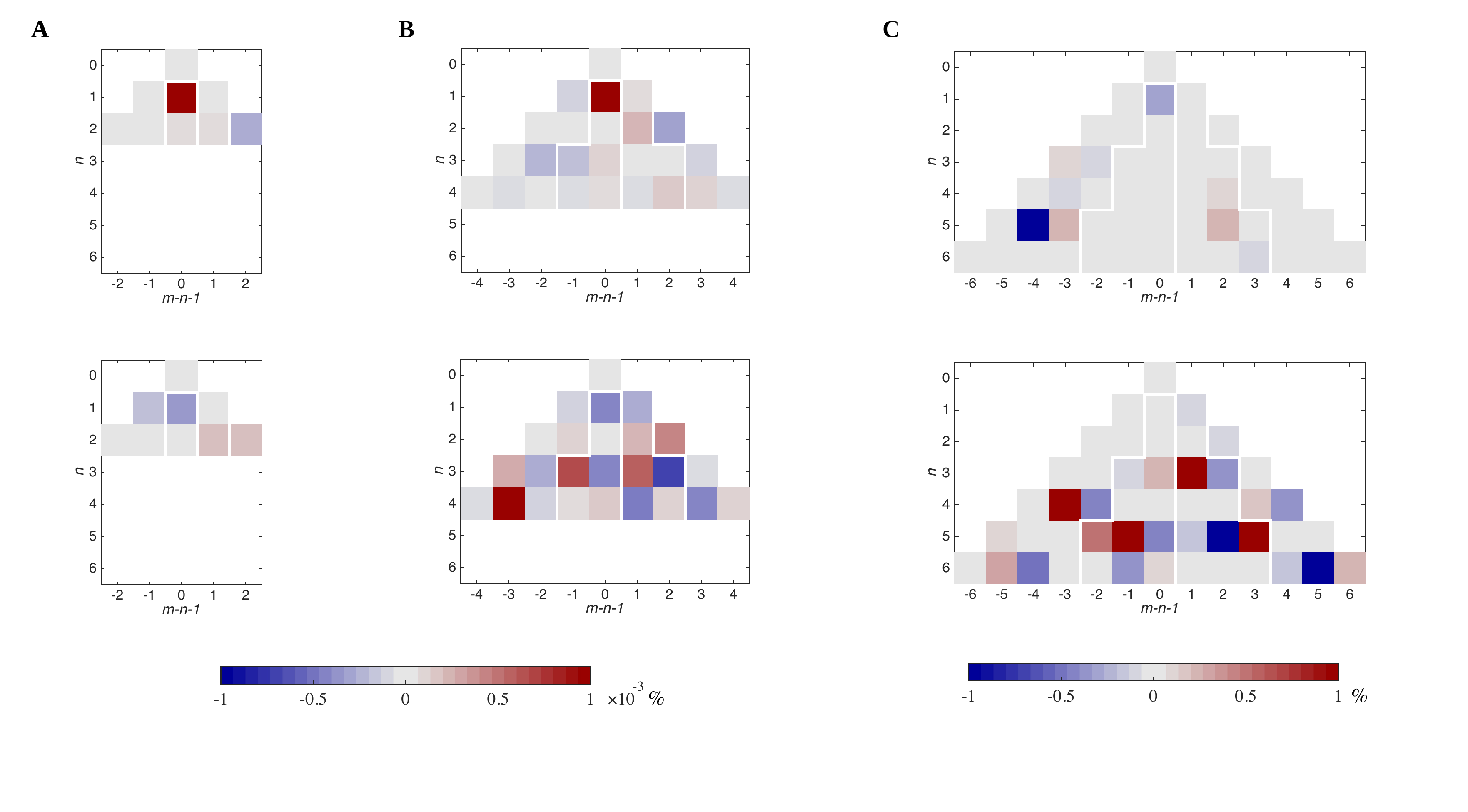}
    \caption{Percentage errors of field coefficients of double EH model up to degree 2 (\textbf{A}), 4 (\textbf{B}), and 6 (\textbf{C}). Results for Lobe 1 are in the first row, those for Lobe 2 in the second. The horizontal axis is defined by the symmetric order, $m-n-1$.} \label{fig:eh-compensate}
\end{figure}

The proposed approach serves two purposes.
First, it offers a further validation of the implemented algorithm for the least-squares solution.
In the simulation setting, where the body density is known, the compensation of omitted (likewise known) coefficients, rather than treating them as noises, should reduce the errors in the estimated EH coefficients.
We performed new solutions for the cases in Figure \ref{fig:eh-hger-res}, where significant estimation errors of tens of percent were present for degree-2 and -4 and all exceeded 100~\% for degree 6.
After compensation, in comparison, the errors were reduced by orders of magnitudes in all cases (Fig. \ref{fig:eh-compensate}).
The absolute errors of the degree-2 and -4 models reach the maxima of 4.5$\times 10^{-3}$~\% and 3.5$\times 10^{-3}$~\% with an RMS of 1.1$\times 10^{-3}$~\% and 5.6$\times 10^{-4}$~\%), respectively.
While still larger, the degree-6 model errors were relieved to the reasonable level of 0.1~\% (maximum: 1.71~\%; RMS: 0.26~\%).

Second, even when the detailed mass distribution is unknown, a uniform body is the first approximation for gravitational field modeling \cite{lhotka2016gravity}.
The GM or the body's bulk average density is also the first parameter to be resolved in the orbit determination.
Any heterogeneity should constitute higher order effects.
Therefore, model compensation assuming homogeneity should offer an improvement to some extent over a plain least-squares solution by (\ref{eq:ls}).

\subsection{Contribution of the neck}
Closely related to the discussion here is the gravitation of the neck, which has so far been neglected.
The neck accounts for 2.4~\% of the total volume.
Treating its density as the unknown parameter, the neck's contribution to the SH field coefficients is evaluated by Eq. (\ref{eq:obs_elem}) with the boundary value computed by the polyhedron method for the \textit{unit} density (or some alternative reference value), say $v^{(0)}$.
The evaluation is performed for all $N$ and $M$.
The resulting $v^{(0)}_{NM}$ can be arbitrarily placed as a separate column into the observation matrix $\mathbf{H}$.
They quantify the impact of the unknown neck density on the SH coefficients.


\section{Solution for arbitrary distributions}
The homogeneous case is suited for the validation of the transformation algorithms and the diagnosis of model errors.
It is a restrictive assumption in reality, even more so for a bilobed body possibly formed from distinct objects.
We investigate next the solutions for some heterogeneous distributions of 67P.

\begin{figure}
    \centering
    \includegraphics[width=\linewidth]{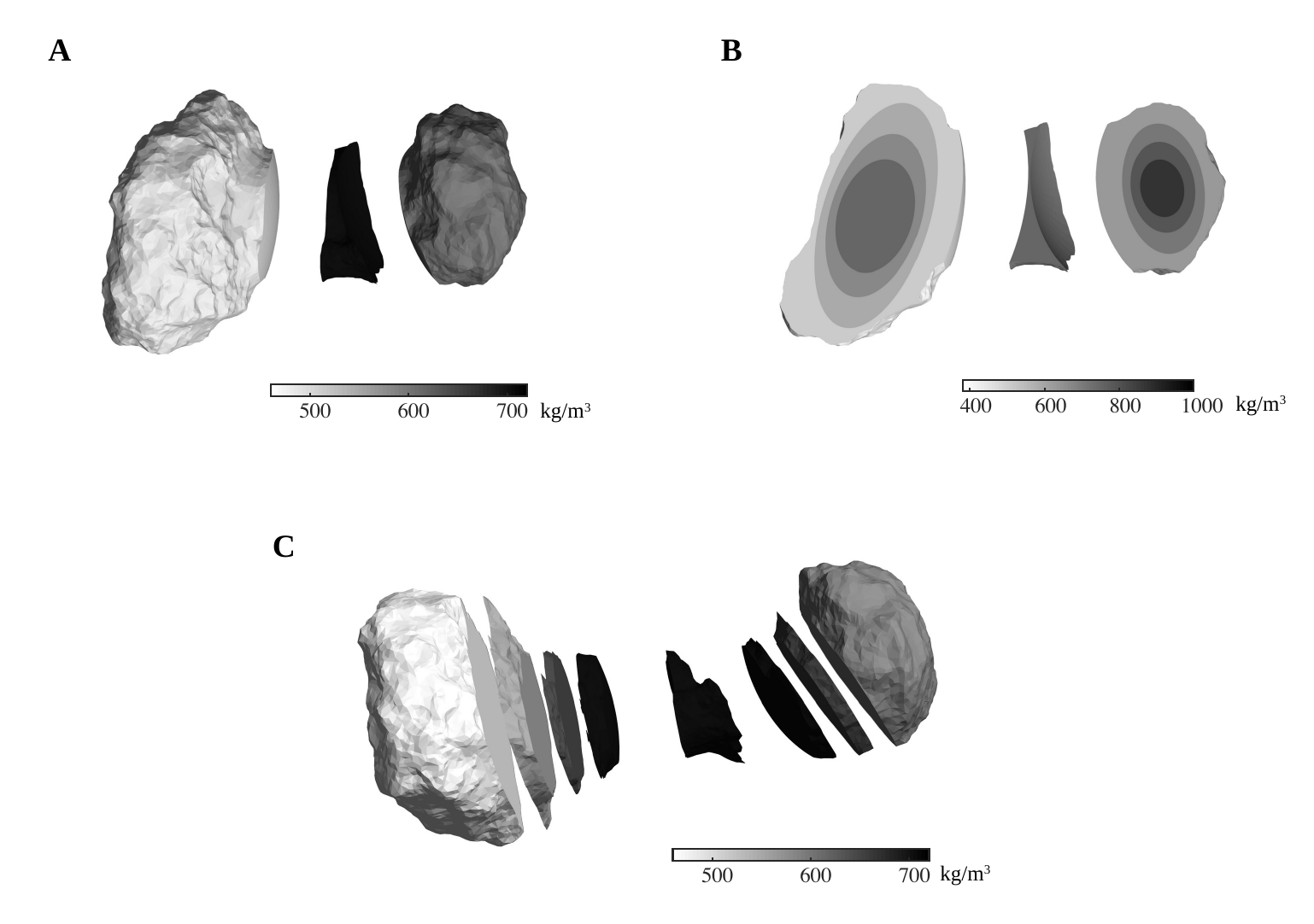}
    \caption{Three simulated heterogeneous distributions of 67P. \textbf{A}. Constant densities for lobes and neck. \textbf{B}. Ellipsoidal stratification, here depicted by their cross sections along global coordinate plane $z=-0.2$~km. \textbf{C}. Densification towards neck. Boundaries between layers (in \textbf{B} and \textbf{C}) specified in Table \ref{tb:geoconfig}.} \label{fg:hetero_illustrate}
\end{figure}

\begin{table}[]
    \centering
    \caption{Density configuration of heterogeneous cases}
    \label{tb:denconfig}
    \begin{tabular}{l l | c | c | c}
    \hline\hline
\multirow{2}{*}{Density} & \multirow{2}{*}{($\mathrm{kg\,m^{-3}}$)} & Lobe 1 & Lobe 2 & \multirow{2}{*}{Neck} \\
 & & \small{(``Body'')} & \small{(``Head'')} & \\
    \hline
       case \i & & 500 & 600 & 700 \\
    \hline
    \multirow{4}{*}{case \i\i}  & Layer 1$^{(\intercal)}$ & 400 & 550 & \multirow{4}{*}{700} \\
    & Layer 2 & 500 & 650 & \\
    & Layer 3 & 600 & 750 & \\
    & Layer 4 & 700 & 850 & \\
    \hline
    \multirow{4}{*}{case \i\i\i}  & Layer 1$^{(*)}$ & 700 & 700 & \multirow{4}{*}{700} \\
    & Layer 2 & 640 & 640 & \\
    & Layer 3 & 550 & 580 & \\
    & Layer 4 & 480 & \, & \\
    \hline
    \multicolumn{5}{l}{($\intercal$) outermost. (*) adjacent to neck.}
    \end{tabular}
\end{table}

\begin{table}[]
    \centering
    \caption{Surfaces of discontinuities of heterogeneous cases \i\i, \i\i\i}
    \label{tb:geoconfig}
    \begin{tabular}{c c | c | c }
    \hline\hline
    \multicolumn{2}{c|}{$f(x,y,z)=1\,^{(*)}$} & Lobe 1 \small{(``Body'')} & Lobe 2 \small{(``Head'')}  \\
    \hline
 \multirow{3}{*}{\parbox{.15\linewidth}{\small{Layer boundary}}}  & 1-2 & $\left(\frac{x}{1.8}\right)^2+\left(\frac{y}{1.3}\right)^2+\left(\frac{z+0.15}{0.7}\right)^2$ & $\left(\frac{x}{1.1}\right)^2+\left(\frac{y}{0.9}\right)^2+\left(\frac{z}{0.6}\right)^2$  \\
 & 2-3 & $\left(\frac{x}{1.3}\right)^2+\left(\frac{y}{1.0}\right)^2+\left(\frac{z+0.15}{0.6}\right)^2$ & $\left(\frac{x}{0.8}\right)^2+\left(\frac{y}{0.7}\right)^2+\left(\frac{z}{0.5}\right)^2$  \\
 &3-4& $\left(\frac{x}{0.9}\right)^2+\left(\frac{y}{0.8}\right)^2+\left(\frac{z+0.15}{0.5}\right)^2$ & $\left(\frac{x}{0.6}\right)^2+\left(\frac{y}{0.5}\right)^2+\left(\frac{z}{0.4}\right)^2$  \\
    \hline
 \multirow{3}{*}{\parbox{.15\linewidth}{\small{Layer boundary}}}  & 1-2 & \multirow{3}{*}{\small{$-0.116x-0.149y+z+\begin{matrix} 0.042\\ 0.144 \\ 0.388 \end{matrix}$}} & \multirow{3}{*}{\small{$0.153x+0.175y+z +\begin{matrix} 1.6\\ 1.4 \\\;\end{matrix}$}} \\
   & 2-3 &  &  \\
   & 3-4 &  &  \\
    \hline
  \multicolumn{4}{l}{(*) $x,y,z$ local coordinates in km.}
    \end{tabular}
\end{table}

\subsection{Case \i: Uniform local densities}
The three-part, constant density contrast between the two lobes and the neck is the simplest form of heterogeneity.
We adopted 500 and 600~$\mathrm{kg/m^3}$ for lobes 1 and 2, respectively.
The neck at 700~$\mathrm{kg/m^3}$ is the densest of all (Tab. \ref{tb:denconfig}), mimicking the possible outcome of merging compaction (Fig. \ref{fg:hetero_illustrate}a).
The global bulk density closely matches that in the homogeneous case.
The EH coefficients of the lobes are given in Table \ref{tb:ehc_heter} in \ref{se:app2}.

We adopted a global SH model up to degree 6 as the truth as well as the observables to derive double EH models up to degree 2.
Without compensation, the estimation errors are comparable to the levels in Figure \ref{fig:eh-hger-res}A.
All coefficients other than $c^{(1)}_{01}$ and $c^{(2)}_{21}$ have errors above 10~\% (Fig. \ref{fg:err_case_i_ii_iii}A).
The errors at degree 1 are the largest with the RMS of 44~\% and 198~\% for lobe 1 and 2, respectively.
The solution for lobe 2 is notably less accurate.
Referring to the test case (Sect.~\ref{ss:test_case}), this can be attributed to the large errors (of 15~\%) in the degree-0 coefficient, which in turn contaminated all of those at degree 1. 

The model compensation proved to be crucial by comparison (Fig. \ref{fg:err_case_i_ii_iii}B).
Even with an erroneous density of 530~$\mathrm{kg\,m^3}$ (and the assumption of homogeneous distributions), the errors were mitigated by accounting for the higher-degree EH coefficients as opposed to omitting them entirely.
With the exception of 163~\% and 20~\% at (1,2) and (2,5), respectively, for lobe 1,\footnote{\ The large errors here were amplified (divided) by the small coefficients, $c^{(1)}_{12},c^{(1)}_{25}$ in comparison (Tab. \ref{tb:ehc_heter}), and do not indicate anomalies.} the errors were reduced to the order of 1~\%.
The lobes' mass were in particular well resolved within an accuracy of 1~\%. 
The error statistics are given in Table \ref{tb:ehc-err_heter}


\subsection{Case \i\i: Stratification}
The lobes themselves may not be uniform.\footnote{\, Neither is the neck any more likely to be uniform; but, its contribution is overall too small to justify a detailed treatment.}
One distinct possibility is the stratification of their interiors. 
For simplicity, we consider a 4-layered structure, where each lobe consists of progressively denser ellipsoidal layers towards the center (Fig. \ref{fg:hetero_illustrate}B).
The layer densities are specified in Table \ref{tb:denconfig} and the boundaries of discontinuity in Table \ref{tb:geoconfig}.

It is interesting that the uncompensated solution has errors similar to that of case~{\i} in both magnitude and pattern (the same is true for case~{\i\i\i}, Fig. \ref{fg:err_case_i_ii_iii}A).
This suggests the substantial errors were predominantly caused by the omitted harmonics, regardless of the mass distributions.
Accordingly, once compensated, the solution became more accurate (Fig. \ref{fg:err_case_i_ii_iii}B, middle column). 
However, because local homogeneity was (erroneously) assumed for the compensation, the solution was expectedly less accurate than in case~{\i} (Fig. \ref{fg:err_case_i_ii_iii}B, left column).
The deterioration is by a few times for Lobe 1 but can reach an order of magnitude in the case of the smaller Lobe 2 (Tab. \ref{tb:ehc-err_heter}).
The mass were distinguished with the respective errors of 0.4~\% and 4.6~\%.
Most estimates at degree 1 suffered errors at the level of 10~\%.
The degree-2 coefficients are overall recovered to the level of 1~\%, notably the oblateness harmonics at (2,1) and (2,2).




\subsection{Case \i\i\i: Gradation}
Another representative scenario is the density increase towards the neck, here treated as 1-D gradation, or planar layering in each lobe (Fig. \ref{fg:hetero_illustrate}, Tab. \ref{tb:geoconfig}). 
The case is relevant because the lobe-neck-lobe partition may not be accurate, i.e., when the neck volume is over- or under-assigned, in which case the density changes do not occur exactly at the designated boundaries.
We assumed that the densities vary from 480 and 580~$\mathrm{kg/m^3}$ at the extremities of Lobe~1 and 2 through four and three increments, respectively, towards the maximum of 700~$\mathrm{kg/m^3}$ at the neck (Tab. \ref{tb:denconfig}).

\begin{figure}
    \centering
    \includegraphics[width=.9 \linewidth]{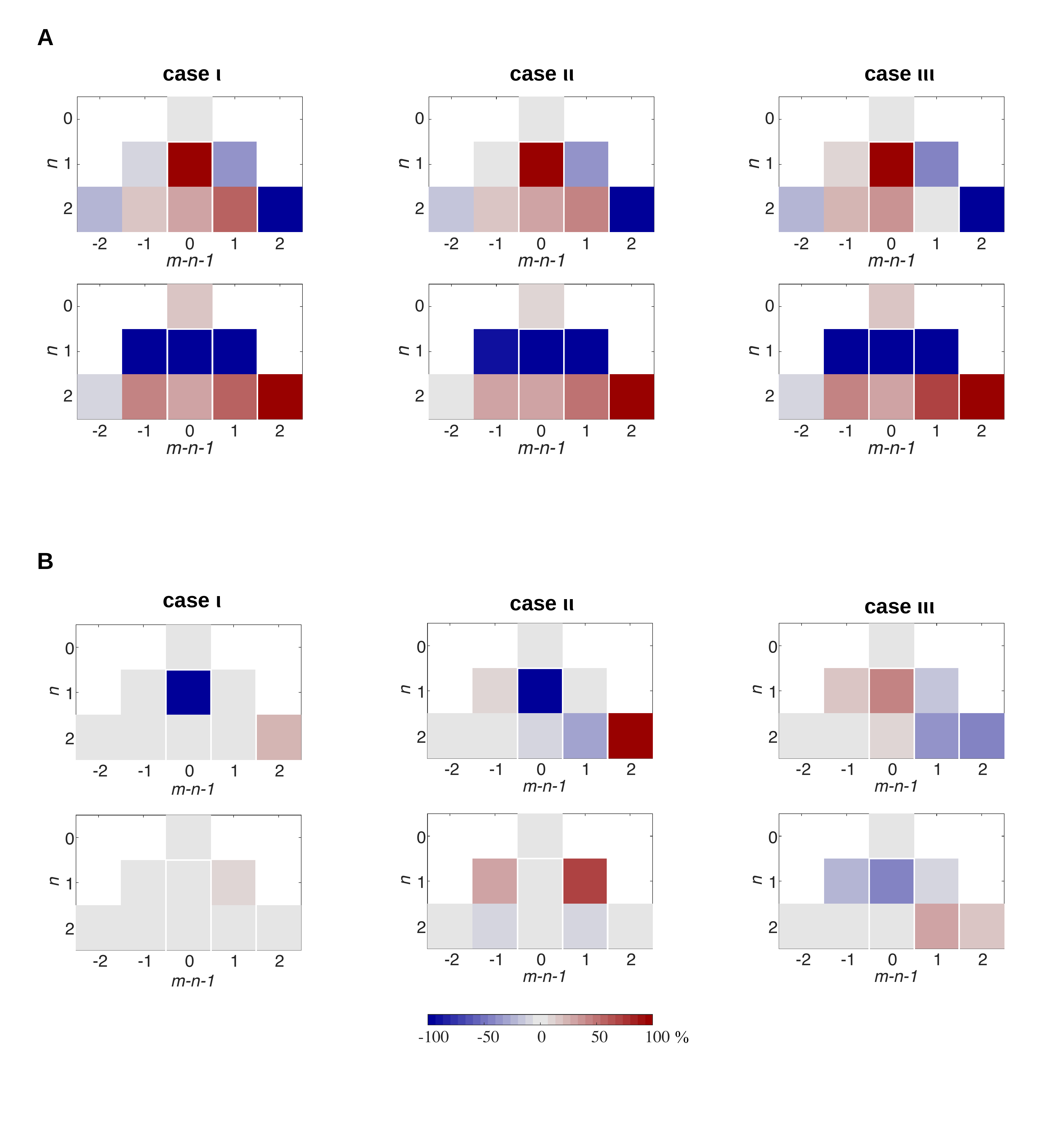}
    \caption{Errors of solved EH coefficient for heterogeneous cases. \textbf{A}. Without compensation. \textbf{B}. With compensation. Results for Lobe 1 are in upper rows in all panels.} \label{fg:err_case_i_ii_iii}
\end{figure}

As in previous cases, model compensation improved the solution decisively (Fig. \ref{fg:err_case_i_ii_iii}, right column).
Larger errors here are associated with Lobe 1, all of which are above 1~\%.
The mass was determined most accurately, with errors of 1.6 and 0.5~\%, followed by the oblateness parameters of a few percent. The errors at the level of 10~\% occur with the others with a maximum of 50~\% (Tab. \ref{tb:ehc-err_heter}).




\section{Discussion \& conclusion}
The double ellipsoidal harmonic-series approach is a natural, viable solution to gravity field modeling of bi-lobed objects.
The series are formulated in the local coordinate systems, and the coefficients distinguishing between the density moments of the individual lobes are conducive to inferring their respective interiors.
The model is strictly valid, i.e., uniformly convergent, in the entire free space outside the ellipsoids fitting the bilobed shape more closely than any other global reference surface.
In case the lobe is closer to a spheroid or a sphere, the spheroidal or spherical harmonic series should be used instead, which would reduce significantly the computational cost.

The model coefficients can be incorporated into the spacecraft orbit determination process in the same way (parameter location) as the SH coefficients.
Out of curiosity we have on the other hand explored an expedient solution from an existing global SH model via decomposition.
In the least-squares procedure, the errors of omitted harmonics beyond the chosen model resolution(s) are expected to be nonzero.
This can be easily appreciated in the solution of the lobes' bulk densities from the degree-1 global model, where the unknown center-of-mass positions of the lobes cause an inevitable bias in the GM estimates and the reconstructed global center-of-mass coordinates.
Rather than leave such errors unaccounted for, we introduced an intuitive approximation that the body would appear homogeneous at shorter wavelengths.
While doing so broke away from the avoidance of artificial assumptions regarding the mass distribution, it yielded a significant improvement on the unaided solution.
The decomposition of a degree-6 SH model into EH models up to degree 2 saw estimation errors largely contained at the level of 10~\% while the mass parameters can be retrieved with an accuracy of about 1~\% or better.

The analysis did not address the solution of the interior mass distribution, for which more detailed constraint or appropriate assumptions is needed.
The complexity of the inverse problem puts it beyond the scope of this prerequisite study, which will be treated in our future work.


\appendix

\section{EH coefficients of degree 1 in relation to center-of-mass Cartesian coordinates} \label{se:app1}
The EH coefficients are related to the body's density moments via \citep{2016CeMDA.125..195H,2017ApJ...850..107H}

\begin{equation}
c_{nm} = \frac{F_{nm}(\rho_0)}{2n+1}\,G \int_\mathcal{M} E_{nm}(\rho) E_{nm}(\mu) E_{nm}(\nu) \mathrm{d}\mathcal{M},
\end{equation}
where the integrand is a solid EH which we denote as $\mathcal{R}_{nm}$. The three Lam{\'e} polynomials of degree~1 are $E_{11}(\mu)=a_{11}\mu$, $E_{12}(\mu)=a_{12}\sqrt{\mu^2-h^2}$, $E_{13}(\mu)=a_{13}\sqrt{k^2-\mu^2}$, where $a_{1m}$ are real (normalized) coefficients. The solid EHs are proportional to the Cartesian coordinates:

\begin{align}
\mathcal{R}_{11} &= (a_{11})^3 \rho \mu \nu = (a_{11})^3 hk\,x \nonumber\\
\mathcal{R}_{12} &= (a_{12})^3 \sqrt{\rho^2-h^2}\,  \sqrt{\mu^2-h^2}\sqrt{h^2-\nu^2} = (a_{12})^3 h \sqrt{k^2-h^2}\, y \nonumber\\
\mathcal{R}_{13} &= (a_{13})^3 \sqrt{\rho^2-k^2}\,  \sqrt{k^2-\mu^2}\sqrt{k^2-\nu^2} = (a_{13})^3 k \sqrt{k^2-h^2} \,z
\end{align}
The EH coefficients are thus given by
\begin{equation}
\left[\begin{matrix}
c_{11} \\ c_{12} \\ c_{13}  \end{matrix}\right] = \frac{1}{3} \left[ \begin{matrix} F_{11}(\rho_0) &(a_{11})^3 hk \\ F_{12}(\rho_0) & (a_{12})^3 h \sqrt{k^2-h^2} \\ F_{13}(\rho_0) & (a_{13})^3 k \sqrt{k^2-h^2} \end{matrix} \right] \,G \int_M \left[\begin{matrix}
    x \\ y \\ z
\end{matrix}\right] \mathrm{d}\mathcal{M}.
\end{equation}
The diagonal elements of $\mathbf{D}^{(i)}$ in Eq. (\ref{eq:eh-com}) correspond respectively to the preceding factors of the GM integral and are therefore distinct from one another.

\section{Ellipsoidal harmonic coefficients and estimation errors for heterogeneous cases} \label{se:app2}

This section contains Tables \ref{tb:ehc_heter} and \ref{tb:ehc-err_heter}.

\begin{table}[]
    \centering
    \caption{True EH coefficients up to degree 2 (Units: $\mathrm{m^2/s^2}$)}
    \label{tb:ehc_heter}
    \begin{tabular}{c | c r | c r | c r | c r }
    \hline\hline
  & \multicolumn{4}{c}{Lobe 1} & \multicolumn{4}{|c}{Lobe 2} \\
    \cline{2-9}
      & \multicolumn{4}{c|}{$n$ $m$ \quad\quad $c^{(1)}_{nm}$} & \multicolumn{4}{c}{$n$ $m$ \quad\quad $c^{(2)}_{nm}$} \\
    \hline
       \multirow{4}{*}{\i} & 0 1 & 0.2410 & 2 1 & -1.049$\times10^{-2}$ & 0 1 & 0.1720 & 2 1 & -7.037$\times10^{-3}$ \\
        & 1 1 & 3.729$\times10^{-3}$ & 2 2 & 1.679$\times10^{-2}$ & 1 1 & -2.375$\times10^{-3}$ & 2 2 & 1.303$\times10^{-2}$ \\
        & 1 2 & 1.067$\times10^{-4}$ & 2 3 & 2.075$\times10^{-3}$ & 1 2 & 1.279$\times10^{-3}$ & 2 3 & -1.852$\times10^{-3}$  \\
        & 1 3 &-1.004$\times10^{-2}$& 2 4 &-1.186$\times10^{-3}$ & 1 3 & -4.827$\times10^{-3}$ & 2 4 & 1.444$\times10^{-3}$  \\
        &    &    & 2 5 & -5.861$\times10^{-4}$ &   &    & 2 5 & 1.322$\times10^{-3}$  \\
    \hline
      \multirow{4}{*}{\i\i} & 0 1 & 0.2361 & 2 1 & -1.117$\times10^{-2}$ & 0 1 & 0.1793 & 2 1 & -7.589$\times10^{-3}$ \\
        & 1 1 & 2.983$\times10^{-3}$ & 2 2 & 1.788$\times10^{-2}$ & 1 1 & -2.177$\times10^{-3}$ & 2 2 & 1.388$\times10^{-2}$ \\
        & 1 2 & 8.538$\times10^{-5}$ & 2 3 & 1.660$\times10^{-3}$ & 1 2 & 1.172$\times10^{-3}$ & 2 3 & -1.698$\times10^{-3}$  \\
        & 1 3 &-1.028$\times10^{-2}$& 2 4 &-9.489$\times10^{-4}$ & 1 3 & -4.425$\times10^{-3}$ & 2 4 & 1.323$\times10^{-3}$  \\
        &    &    & 2 5 & -4.688$\times10^{-4}$ &   &    & 2 5 & 1.212$\times10^{-3}$  \\ 
    \hline
       \multirow{4}{*}{\i\i\i} & 0 1 & 0.2411 & 2 1 & -1.079$\times10^{-2}$ & 0 1 & 0.1727 & 2 1 & -7.118$\times10^{-3}$ \\
        & 1 1 & 4.843$\times10^{-3}$ & 2 2 & 1.787$\times10^{-2}$ & 1 1 & -3.043$\times10^{-3}$ & 2 2 & 1.362$\times10^{-2}$ \\
        & 1 2 & 1.357$\times10^{-3}$ & 2 3 & 2.274$\times10^{-3}$ & 1 2 & 6.853$\times10^{-4}$ & 2 3 & -1.753$\times10^{-3}$  \\
        & 1 3 &-1.237$\times10^{-2}$& 2 4 &-1.893$\times10^{-3}$ & 1 3 & -6.498$\times10^{-3}$ & 2 4 & 1.809$\times10^{-3}$  \\
        &    &    & 2 5 & -1.405$\times10^{-3}$ &   &    & 2 5 & 1.606$\times10^{-3}$  \\
    \hline
    \end{tabular}
\end{table}

\begin{table}[]
    \centering
    \caption{Errors (\%) of estimated EH coefficients}
    \label{tb:ehc-err_heter}
    \begin{tabular}{c | c r | c r | c r | c r }
    \hline\hline
  & \multicolumn{4}{c}{Lobe 1} & \multicolumn{4}{|c}{Lobe 2} \\
    \cline{2-9}
      & \multicolumn{4}{c|}{$n$ $m$ \quad\quad $c^{(1)}_{nm}$} & \multicolumn{4}{c}{$n$ $m$ \quad\quad $c^{(2)}_{nm}$} \\
    \hline
       \multirow{4}{*}{\i} & 0 1 & -0.2393 & 2 1 & 1.26 & 0 1 & -0.6401 & 2 1 & 0.1744 \\
        & 1 1 & 0.376 & 2 2 & -0.8943 & 1 1 & 1.406 & 2 2 & -0.5933 \\
        & 1 2 & -163.4 & 2 3 & -2.058 & 1 2 & 5.5 & 2 3 & 1.288  \\
        & 1 3 & 1.931 & 2 4 & -3.23 & 1 3 & 6.921 & 2 4 & 4.257  \\
        &    &    & 2 5 & 20.16 &   &    & 2 5 & 0.8726  \\
    \hline
       \multirow{4}{*}{\i\i} & 0 1 & -0.4141 & 2 1 & 3.2 & 0 1 & -4.66 & 2 1 & 4.375 \\
        & 1 1 & 10.38 & 2 2 & -0.8383 & 1 1 & 26.92 & 2 2 & -9.699 \\
        & 1 2 & -732.7 & 2 3 & -9.348 & 1 2 & 5.542 & 2 3 & 2.149  \\
        & 1 3 & 2.08 & 2 4 & -27.92 & 1 3 & 70.06 & 2 4 & -6.714  \\
        &    &    & 2 5 & 96.55 &   &    & 2 5 & -0.7541  \\ 
    \hline
       \multirow{4}{*}{\i\i\i} & 0 1 & 1.649 & 2 1 & -2.9 & 0 1 & 0.5463 & 2 1 & -0.85 \\
        & 1 1 & 16.66 & 2 2 & 6.115 & 1 1 &-25.07 & 2 2 & 3.953 \\
        & 1 2 & 46.56 & 2 3 & 7.223 & 1 2 & -44.86 & 2 3 & 0.8824  \\
        & 1 3 & -13.4 & 2 4 & -39.28 & 1 3 & -7.597 & 2 4 & 29.04  \\
        &    &    & 2 5 & -42.29 &   &    & 2 5 & 19.33  \\ 
    \hline
    \end{tabular}
\end{table}

\vspace{10 mm}
\noindent \textbf{Acknowledgement}\\
Hu acknowledges financial support from the German Research Foundation under project ``Geodesy of small bodies: From gravitation to interior structure'' (no. 500329796).
\\


 \bibliographystyle{elsarticle-num} 
 \bibliography{reference}

\begin{thebibliography}{10}
\expandafter\ifx\csname url\endcsname\relax
  \def\url#1{\texttt{#1}}\fi
\expandafter\ifx\csname urlprefix\endcsname\relax\def\urlprefix{URL }\fi
\expandafter\ifx\csname href\endcsname\relax
  \def\href#1#2{#2} \def\path#1{#1}\fi

\bibitem{2015Sci...347a1044S}
H.~{Sierks}, C.~{Barbieri}, P.~L. {Lamy}, R.~{Rodrigo}, D.~{Koschny}, H.~{Rickman}, H.~U. {Keller}, J.~{Agarwal}, M.~F. {A'Hearn}, F.~{Angrilli}, A.-T. {Auger}, M.~A. {Barucci}, J.-L. {Bertaux}, I.~{Bertini}, S.~{Besse}, D.~{Bodewits}, C.~{Capanna}, G.~{Cremonese}, V.~{Da Deppo}, B.~{Davidsson}, S.~{Debei}, M.~{De Cecco}, F.~{Ferri}, S.~{Fornasier}, M.~{Fulle}, R.~{Gaskell}, L.~{Giacomini}, O.~{Groussin}, P.~{Gutierrez-Marques}, P.~J. {Guti{\'e}rrez}, C.~{G{\"u}ttler}, N.~{Hoekzema}, S.~F. {Hviid}, W.-H. {Ip}, L.~{Jorda}, J.~{Knollenberg}, G.~{Kovacs}, J.~R. {Kramm}, E.~{K{\"u}hrt}, M.~{K{\"u}ppers}, F.~{La Forgia}, L.~M. {Lara}, M.~{Lazzarin}, C.~{Leyrat}, J.~J. {Lopez Moreno}, S.~{Magrin}, S.~{Marchi}, F.~{Marzari}, M.~{Massironi}, H.~{Michalik}, R.~{Moissl}, S.~{Mottola}, G.~{Naletto}, N.~{Oklay}, M.~{Pajola}, M.~{Pertile}, F.~{Preusker}, L.~{Sabau}, F.~{Scholten}, C.~{Snodgrass}, N.~{Thomas}, C.~{Tubiana}, J.-B. {Vincent}, K.-P. {Wenzel}, M.~{Zaccariotto}, M.~{P{\"a}tzold}, {On the nucleus structure and
  activity of comet 67P/Churyumov-Gerasimenko}, Science 347~(6220) (2015) aaa1044.
\newblock \href {https://doi.org/10.1126/science.aaa1044} {\path{doi:10.1126/science.aaa1044}}.

\bibitem{2006Sci...312.1330F}
A.~{Fujiwara}, J.~{Kawaguchi}, D.~K. {Yeomans}, M.~{Abe}, T.~{Mukai}, T.~{Okada}, J.~{Saito}, H.~{Yano}, M.~{Yoshikawa}, D.~J. {Scheeres}, O.~{Barnouin-Jha}, A.~F. {Cheng}, H.~{Demura}, R.~W. {Gaskell}, N.~{Hirata}, H.~{Ikeda}, T.~{Kominato}, H.~{Miyamoto}, A.~M. {Nakamura}, R.~{Nakamura}, S.~{Sasaki}, K.~{Uesugi}, {The Rubble-Pile Asteroid Itokawa as Observed by Hayabusa}, Science 312~(5778) (2006) 1330--1334.
\newblock \href {https://doi.org/10.1126/science.1125841} {\path{doi:10.1126/science.1125841}}.

\bibitem{2013NatSR...3E3411H}
J.~{Huang}, J.~{Ji}, P.~{Ye}, X.~{Wang}, J.~{Yan}, L.~{Meng}, S.~{Wang}, C.~{Li}, Y.~{Li}, D.~{Qiao}, W.~{Zhao}, Y.~{Zhao}, T.~{Zhang}, P.~{Liu}, Y.~{Jiang}, W.~{Rao}, S.~{Li}, C.~{Huang}, W.-H. {Ip}, S.~{Hu}, M.~{Zhu}, L.~{Yu}, Y.~{Zou}, X.~{Tang}, J.~{Li}, H.~{Zhao}, H.~{Huang}, X.~{Jiang}, J.~{Bai}, {The Ginger-shaped Asteroid 4179 Toutatis: New Observations from a Successful Flyby of Chang'e-2}, Scientific Reports 3 (2013) 3411.
\newblock \href {http://arxiv.org/abs/1312.4329} {\path{arXiv:1312.4329}}, \href {https://doi.org/10.1038/srep03411} {\path{doi:10.1038/srep03411}}.

\bibitem{stern2019initial}
S.~Stern, H.~Weaver, J.~Spencer, C.~Olkin, G.~Gladstone, W.~Grundy, J.~Moore, D.~Cruikshank, H.~Elliott, W.~McKinnon, et~al., Initial results from the new horizons exploration of 2014 mu69, a small kuiper belt object, Science 364~(6441) (2019) eaaw9771.

\bibitem{levison2024contact}
H.~F. Levison, S.~Marchi, K.~S. Noll, J.~R. Spencer, T.~S. Statler, J.~F. Bell~III, E.~B. Bierhaus, R.~Binzel, W.~F. Bottke, D.~Britt, et~al., A contact binary satellite of the asteroid (152830) dinkinesh, Nature 629~(8014) (2024) 1015--1020.

\bibitem{2015aste.book..355M}
J.~L. {Margot}, P.~{Pravec}, P.~{Taylor}, B.~{Carry}, S.~{Jacobson}, {Asteroid Systems: Binaries, Triples, and Pairs}, in: Asteroids IV, 2015, pp. 355--374.
\newblock \href {https://doi.org/10.2458/azu\_uapress\_9780816532131-ch019} {\path{doi:10.2458/azu\_uapress\_9780816532131-ch019}}.

\bibitem{massironi2015two}
M.~Massironi, E.~Simioni, F.~Marzari, G.~Cremonese, L.~Giacomini, M.~Pajola, L.~Jorda, G.~Naletto, S.~Lowry, M.~R. El-Maarry, et~al., Two independent and primitive envelopes of the bilobate nucleus of comet 67p, Nature 526~(7573) (2015) 402--405.

\bibitem{jutzi2017formation}
M.~Jutzi, W.~Benz, Formation of bi-lobed shapes by sub-catastrophic collisions-a late origin of comet 67p’s structure, Astronomy \& Astrophysics 597 (2017) A62.

\bibitem{mckinnon2020solar}
W.~McKinnon, D.~Richardson, J.~Marohnic, J.~Keane, W.~Grundy, D.~Hamilton, D.~Nesvorn{\`y}, O.~Umurhan, T.~Lauer, K.~Singer, et~al., The solar nebula origin of (486958) arrokoth, a primordial contact binary in the kuiper belt, Science 367~(6481) (2020) eaay6620.

\bibitem{MG2000}
G.~{Montenbruck}, E.~{Gill}, Satellite Orbits, Springer-Verlag, Berlin Heidelberg, 2000.

\bibitem{TAPLEY2004xi}
B.~{Tapley}, B.~{Schutz}, G.~{Born}, \href{https://www.sciencedirect.com/science/article/pii/B9780126836301500199}{Statistical Orbit Determination}, Academic Press, Burlington, 2004.
\newblock \href {https://doi.org/https://doi.org/10.1016/B978-012683630-1/50019-9} {\path{doi:https://doi.org/10.1016/B978-012683630-1/50019-9}}.
\newline\urlprefix\url{https://www.sciencedirect.com/science/article/pii/B9780126836301500199}

\bibitem{paul1974gravity}
M.~Paul, The gravity effect of a homogeneous polyhedron for three-dimensional interpretation, Pure and Applied Geophysics 112 (1974) 553--561.

\bibitem{pohanka1998optimum}
V.~Poh{\'a}nka, Optimum expression for computation of the gravity field of a polyhedral body with linearly increasing density, Geophysical Prospecting 46~(4) (1998) 391--404.

\bibitem{balmino1994gravitational}
G.~Balmino, Gravitational potential harmonics from the shape of an homogeneous body, Celestial Mechanics and Dynamical Astronomy 60 (1994) 331--364.

\bibitem{russell2012global}
R.~P. Russell, N.~Arora, Global point mascon models for simple, accurate, and parallel geopotential computation, Journal of Guidance, Control, and Dynamics 35~(5) (2012) 1568--1581.

\bibitem{1996CeMDA..65..313W}
R.~A. {Werner}, D.~J. {Scheeres}, {Exterior gravitation of a polyhedron derived and compared with harmonic and mascon gravitation representations of asteroid 4769 Castalia}, Celestial Mechanics and Dynamical Astronomy 65~(3) (1996) 313--344.
\newblock \href {https://doi.org/10.1007/BF00053511} {\path{doi:10.1007/BF00053511}}.

\bibitem{chen2019spherical}
C.~Chen, Y.~Ouyang, S.~Bian, Spherical harmonic expansions for the gravitational field of a polyhedral body with polynomial density contrast, Surveys in Geophysics 40 (2019) 197--246.

\bibitem{park2010estimating}
R.~S. Park, R.~A. Werner, S.~Bhaskaran, Estimating small-body gravity field from shape model and navigation data, Journal of guidance, control, and dynamics 33~(1) (2010) 212--221.

\bibitem{tricarico2013global}
P.~Tricarico, Global gravity inversion of bodies with arbitrary shape, Geophysical Journal International 195~(1) (2013) 260--275.

\bibitem{takahashi2014morphology}
Y.~Takahashi, D.~Scheeres, Morphology driven density distribution estimation for small bodies, Icarus 233 (2014) 179--193.

\bibitem{andert2015gravity}
T.~Andert, J.-P. Barriot, M.~Paetzold, L.~Sichoix, S.~Tellmann, B.~H{\"a}usler, The gravity field of comet 67 p/churyumov-gerasimenko expressed in bispherical harmonics, in: AGU Fall Meeting Abstracts, Vol. 2015, 2015, pp. P31E--2109.

\bibitem{jeffery1912form}
G.~B. Jeffery, On a form of the solution of laplace's equation suitable for problems relating to two spheres, Proceedings of the Royal Society of London. Series A, Containing Papers of a Mathematical and Physical Character 87~(593) (1912) 109--120.

\bibitem{zeng2015study}
X.~Zeng, F.~Jiang, J.~Li, H.~Baoyin, Study on the connection between the rotating mass dipole and natural elongated bodies, Astrophysics and Space Science 356 (2015) 29--42.

\bibitem{burov2019approximation}
A.~A. Burov, A.~D. Guerman, E.~A. Nikonova, V.~I. Nikonov, Approximation for attraction field of irregular celestial bodies using four massive points, Acta Astronautica 157 (2019) 225--232.

\bibitem{scheeres1994dynamics}
D.~J. Scheeres, Dynamics about uniformly rotating triaxial ellipsoids: applications to asteroids, Icarus 110~(2) (1994) 225--238.

\bibitem{wei2020hybrid}
B.~Wei, H.~Shang, D.~Qiao, Hybrid model of gravitational fields around small bodies for efficient trajectory propagations, Journal of Guidance, Control, and Dynamics 43~(2) (2020) 232--249.

\bibitem{2001CeMDA..79..235R}
R.~{Garmier}, J.-P. {Barriot}, {Ellipsoidal Harmonic expansions of the gravitational potential: Theory and application}, Celestial Mechanics and Dynamical Astronomy 79~(4) (2001) 235--275.
\newblock \href {https://doi.org/10.1023/A:1017555515763} {\path{doi:10.1023/A:1017555515763}}.

\bibitem{2016CeMDA.125..195H}
X.~{Hu}, {The exact transformation from spherical harmonic to ellipsoidal harmonic coefficients for gravitational field modeling}, Celestial Mechanics and Dynamical Astronomy 125~(2) (2016) 195--222.
\newblock \href {https://doi.org/10.1007/s10569-016-9678-z} {\path{doi:10.1007/s10569-016-9678-z}}.

\bibitem{2002GeoRL..29.1231G}
R.~{Garmier}, J.-P. {Barriot}, A.~S. {Konopliv}, D.~K. {Yeomans}, {Modeling of the Eros gravity field as an ellipsoidal harmonic expansion from the NEAR Doppler tracking data}, Geophysical Research Letters 29~(8) (2002) 1231.
\newblock \href {https://doi.org/10.1029/2001GL013768} {\path{doi:10.1029/2001GL013768}}.

\bibitem{park2014gravity}
R.~Park, A.~Konopliv, S.~Asmar, B.~Bills, R.~Gaskell, C.~Raymond, D.~Smith, M.~Toplis, M.~Zuber, Gravity field expansion in ellipsoidal harmonic and polyhedral internal representations applied to vesta, Icarus 240 (2014) 118--132.

\bibitem{2016Icar..277..257J}
L.~{Jorda}, R.~{Gaskell}, C.~{Capanna}, S.~{Hviid}, P.~{Lamy}, J.~{{\v{D}}urech}, G.~{Faury}, O.~{Groussin}, P.~{Guti{\'e}rrez}, C.~{Jackman}, S.~J. {Keihm}, H.~U. {Keller}, J.~{Knollenberg}, E.~{K{\"u}hrt}, S.~{Marchi}, S.~{Mottola}, E.~{Palmer}, F.~P. {Schloerb}, H.~{Sierks}, J.~B. {Vincent}, M.~F. {A'Hearn}, C.~{Barbieri}, R.~{Rodrigo}, D.~{Koschny}, H.~{Rickman}, M.~A. {Barucci}, J.~L. {Bertaux}, I.~{Bertini}, G.~{Cremonese}, V.~{Da Deppo}, B.~{Davidsson}, S.~{Debei}, M.~{De Cecco}, S.~{Fornasier}, M.~{Fulle}, C.~{G{\"u}ttler}, W.~H. {Ip}, J.~R. {Kramm}, M.~{K{\"u}ppers}, L.~M. {Lara}, M.~{Lazzarin}, J.~J. {Lopez Moreno}, F.~{Marzari}, G.~{Naletto}, N.~{Oklay}, N.~{Thomas}, C.~{Tubiana}, K.~P. {Wenzel}, {The global shape, density and rotation of Comet 67P/Churyumov-Gerasimenko from preperihelion Rosetta/OSIRIS observations}, Icarus 277 (2016) 257--278.
\newblock \href {https://doi.org/10.1016/j.icarus.2016.05.002} {\path{doi:10.1016/j.icarus.2016.05.002}}.

\bibitem{HM1967}
W.~A. {Heiskanen}, H.~{Moritz}, {Physical Geodesy}, W.H. Freeman and Company San Francisco, 1967.

\bibitem{patzold2016homogeneous}
M.~P{\"a}tzold, T.~Andert, M.~Hahn, S.~Asmar, J.-P. Barriot, M.~Bird, B.~H{\"a}usler, K.~Peter, S.~Tellmann, E.~Gr{\"u}n, et~al., A homogeneous nucleus for comet 67p/churyumov--gerasimenko from its gravity field, Nature 530~(7588) (2016) 63--65.

\bibitem{patzold2019nucleus}
M.~P{\"a}tzold, T.~P. Andert, M.~Hahn, J.-P. Barriot, S.~W. Asmar, B.~H{\"a}usler, M.~K. Bird, S.~Tellmann, J.~Oschlisniok, K.~Peter, The nucleus of comet 67p/churyumov--gerasimenko--part i: The global view--nucleus mass, mass-loss, porosity, and implications, Monthly Notices of the Royal Astronomical Society 483~(2) (2019) 2337--2346.

\bibitem{2015JGeod..89..159H}
X.~{Hu}, C.~{Jekeli}, {A numerical comparison of spherical, spheroidal and ellipsoidal harmonic gravitational field models for small non-spherical bodies: examples for the Martian moons}, Journal of Geodesy 89~(2) (2015) 159--177.
\newblock \href {https://doi.org/10.1007/s00190-014-0769-x} {\path{doi:10.1007/s00190-014-0769-x}}.

\bibitem{2016JGRE..121..497R}
S.~{Reimond}, O.~{Baur}, {Spheroidal and ellipsoidal harmonic expansions of the gravitational potential of small Solar System bodies. Case study: Comet 67P/Churyumov-Gerasimenko}, Journal of Geophysical Research (Planets) 121~(3) (2016) 497--515.
\newblock \href {http://arxiv.org/abs/1610.06491} {\path{arXiv:1610.06491}}, \href {https://doi.org/10.1002/2015JE004965} {\path{doi:10.1002/2015JE004965}}.

\bibitem{lhotka2016gravity}
C.~Lhotka, S.~Reimond, J.~Souchay, O.~Baur, Gravity field and solar component of the precession rate and nutation coefficients of comet 67p/churyumov--gerasimenko, Monthly Notices of the Royal Astronomical Society 455~(4) (2016) 3588--3596.

\bibitem{2017ApJ...850..107H}
X.~{Hu}, {Normal Gravity Fields and Equipotential Ellipsoids of Small Objects in the Solar System: A Closed-form Solution in Ellipsoidal Harmonics up to the Second Degree}, The Astrophysical Journal 850~(1) (2017) 107.
\newblock \href {https://doi.org/10.3847/1538-4357/aa9222} {\path{doi:10.3847/1538-4357/aa9222}}.

\end{thebibliography}





\end{document}